\documentclass[sigconf]{acmart}
\AtBeginDocument{%
  }

\newcommand{\tableResultsEnergyAccuracy}{
\begin{table*}[t!]
\centering
\scriptsize
\begin{subtable}{\textwidth}
\centering
\begin{tabular}{|l|ccccccccc|ccccccclc|}
\hline
\multicolumn{1}{|c|}{\multirow{3}{*}{}} & \multicolumn{9}{c|}{\textbf{HumanEval}}                                                                                                                                                                                                                                                                                                                                                                        & \multicolumn{9}{c|}{\textbf{MBPP}}                                                                                                                                                                                                                                                                                                                                                                                                           \\ \cline{2-19} 
\multicolumn{1}{|c|}{}                  & \multicolumn{3}{c|}{\textbf{GPU Energy}}                                                                                                       & \multicolumn{3}{c|}{\textbf{Output}}                                                                                                       & \multicolumn{3}{c|}{\textbf{Accuracy}}                                                                               & \multicolumn{3}{c|}{\textbf{GPU Energy}}                                                                                                       & \multicolumn{3}{c|}{\textbf{Output}}                                                                                                                       & \multicolumn{3}{c|}{\textbf{Accuracy}}                                                                                             \\ \cline{2-19} 
\multicolumn{1}{|c|}{}                  & \multicolumn{1}{c|}{\textbf{Baseline}} & \multicolumn{1}{c|}{\textbf{BS}} & \multicolumn{1}{c|}{\textbf{$\Delta$}}                         & \multicolumn{1}{c|}{\textbf{Baseline}} & \multicolumn{1}{c|}{\textbf{BS}} & \multicolumn{1}{c|}{\textbf{$\Delta$}}                         & \multicolumn{1}{c|}{\textbf{Baseline}} & \multicolumn{1}{c|}{\textbf{BS}} & \textbf{$\Delta$}                        & \multicolumn{1}{c|}{\textbf{Baseline}} & \multicolumn{1}{c|}{\textbf{BS}} & \multicolumn{1}{c|}{\textbf{$\Delta$}}                         & \multicolumn{1}{c|}{\textbf{Baseline}} & \multicolumn{1}{c|}{\textbf{BS}} & \multicolumn{1}{c|}{\textbf{$\Delta$}}                                         & \multicolumn{1}{c|}{\textbf{Baseline}} & \multicolumn{1}{c|}{\textbf{BS}} & \textbf{$\Delta$}                                      \\ \hline
\textbf{CodeLlama-7B}                   & \multicolumn{1}{c|}{3242}              & \multicolumn{1}{c|}{2578}        & \multicolumn{1}{c|}{21\%$\textcolor{ForestGreen}{\downarrow}$} & \multicolumn{1}{c|}{600}               & \multicolumn{1}{c|}{461}         & \multicolumn{1}{c|}{\textbf{23\%}$\textcolor{ForestGreen}{\downarrow}$} & \multicolumn{1}{c|}{33\%}              & \multicolumn{1}{c|}{33\%}        & 0\%                                      & \multicolumn{1}{c|}{4754}              & \multicolumn{1}{c|}{2889}        & \multicolumn{1}{c|}{39\%$\textcolor{ForestGreen}{\downarrow}$} & \multicolumn{1}{c|}{881}               & \multicolumn{1}{c|}{517}         & \multicolumn{1}{c|}{\textbf{41\%}$\textcolor{ForestGreen}{\downarrow}$}                 & \multicolumn{1}{c|}{43\%}              & \multicolumn{1}{l|}{43\%}        & 0\%                                                    \\ \hline
\textbf{Qwen2.5-Coder-7B}               & \multicolumn{1}{c|}{625}               & \multicolumn{1}{c|}{672}         & \multicolumn{1}{c|}{8\%$\textcolor{red}{\uparrow}$}            & \multicolumn{1}{c|}{120}               & \multicolumn{1}{c|}{118}         & \multicolumn{1}{c|}{2\%$\textcolor{ForestGreen}{\downarrow}$}  & \multicolumn{1}{c|}{70\%}              & \multicolumn{1}{c|}{70\%}        & 0\%                                      & \multicolumn{1}{c|}{241}               & \multicolumn{1}{c|}{284}         & \multicolumn{1}{c|}{18\%$\textcolor{red}{\uparrow}$}           & \multicolumn{1}{c|}{47}                & \multicolumn{1}{c|}{47}          & \multicolumn{1}{c|}{0\%}                                                       & \multicolumn{1}{c|}{64\%}              & \multicolumn{1}{l|}{64\%}        & 0\%                                                    \\ \hline
\textbf{Deepseek-6.7B}                  & \multicolumn{1}{c|}{5347}              & \multicolumn{1}{c|}{2216}        & \multicolumn{1}{c|}{\textbf{58\%}$\textcolor{ForestGreen}{\downarrow}$} & \multicolumn{1}{c|}{983}               & \multicolumn{1}{c|}{397}         & \multicolumn{1}{c|}{\textbf{60\%}$\textcolor{ForestGreen}{\downarrow}$} & \multicolumn{1}{c|}{65\%}              & \multicolumn{1}{c|}{65\%}        & 0\%                                      & \multicolumn{1}{c|}{5357}              & \multicolumn{1}{c|}{2449}        & \multicolumn{1}{c|}{\textbf{54\%}$\textcolor{ForestGreen}{\downarrow}$} & \multicolumn{1}{c|}{1000}              & \multicolumn{1}{c|}{440}         & \multicolumn{1}{c|}{\textbf{56\%}$\textcolor{ForestGreen}{\downarrow}$} & \multicolumn{1}{c|}{58\%}              & \multicolumn{1}{l|}{58\%}        & 0\%     \\ \hline
\textbf{CodeGemma-7B}                   & \multicolumn{1}{c|}{640}               & \multicolumn{1}{c|}{695}         & \multicolumn{1}{c|}{9\%$\textcolor{red}{\uparrow}$}            & \multicolumn{1}{c|}{99}                & \multicolumn{1}{c|}{93}          & \multicolumn{1}{c|}{6\%$\textcolor{ForestGreen}{\downarrow}$}  & \multicolumn{1}{c|}{54\%}              & \multicolumn{1}{c|}{54\%}        & 0\%                                      & \multicolumn{1}{c|}{880}               & \multicolumn{1}{c|}{823}         & \multicolumn{1}{c|}{7\%$\textcolor{ForestGreen}{\downarrow}$}  & \multicolumn{1}{c|}{130}               & \multicolumn{1}{c|}{108}         & \multicolumn{1}{c|}{17\%$\textcolor{ForestGreen}{\downarrow}$}                 & \multicolumn{1}{c|}{50\%}              & \multicolumn{1}{l|}{50\%}        & 0\%                                                    \\ \hline
\textbf{CodeQwen1.5-7B}                 & \multicolumn{1}{c|}{1030}              & \multicolumn{1}{c|}{972}         & \multicolumn{1}{c|}{6\%$\textcolor{ForestGreen}{\downarrow}$}  & \multicolumn{1}{c|}{199}               & \multicolumn{1}{c|}{175}         & \multicolumn{1}{c|}{12\%$\textcolor{ForestGreen}{\downarrow}$} & \multicolumn{1}{c|}{52\%}              & \multicolumn{1}{c|}{52\%}        & 0\%                                      & \multicolumn{1}{c|}{839}               & \multicolumn{1}{c|}{795}         & \multicolumn{1}{c|}{\textbf{5\%}$\textcolor{ForestGreen}{\downarrow}$}  & \multicolumn{1}{c|}{163}               & \multicolumn{1}{c|}{142}         & \multicolumn{1}{c|}{\textbf{13\%}$\textcolor{ForestGreen}{\downarrow}$} & \multicolumn{1}{c|}{54\%}              & \multicolumn{1}{l|}{54\%}        & 0\%                                                    \\ \hline
\textbf{NextCoder-7B}                   & \multicolumn{1}{c|}{2045}              & \multicolumn{1}{c|}{1919}        & \multicolumn{1}{c|}{6\%$\textcolor{ForestGreen}{\downarrow}$}  & \multicolumn{1}{c|}{392}               & \multicolumn{1}{c|}{359}         & \multicolumn{1}{c|}{8\%$\textcolor{ForestGreen}{\downarrow}$}  & \multicolumn{1}{c|}{41\%}              & \multicolumn{1}{c|}{41\%}        & 0\%                                      & \multicolumn{1}{c|}{1519}              & \multicolumn{1}{c|}{763}         & \multicolumn{1}{c|}{\textbf{50\%}$\textcolor{ForestGreen}{\downarrow}$} & \multicolumn{1}{c|}{290}               & \multicolumn{1}{c|}{135}         & \multicolumn{1}{c|}{\textbf{54\%}$\textcolor{ForestGreen}{\downarrow}$}                 & \multicolumn{1}{c|}{54\%}              & \multicolumn{1}{l|}{54\%}        & 0\%                                                    \\ \hline
\textbf{Phi3.5-4B}                      & \multicolumn{1}{c|}{787}               & \multicolumn{1}{c|}{621}         & \multicolumn{1}{c|}{\textbf{21\%}$\textcolor{ForestGreen}{\downarrow}$} & \multicolumn{1}{c|}{189}               & \multicolumn{1}{c|}{136}         & \multicolumn{1}{c|}{\textbf{28\%}$\textcolor{ForestGreen}{\downarrow}$} & \multicolumn{1}{c|}{70\%}              & \multicolumn{1}{c|}{70\%}        & 0\%                                      & \multicolumn{1}{c|}{723}               & \multicolumn{1}{c|}{412}         & \multicolumn{1}{c|}{\textbf{43\%}$\textcolor{ForestGreen}{\downarrow}$} & \multicolumn{1}{c|}{167}               & \multicolumn{1}{c|}{92}          & \multicolumn{1}{c|}{\textbf{45\%}$\textcolor{ForestGreen}{\downarrow}$}                 & \multicolumn{1}{c|}{54\%}              & \multicolumn{1}{l|}{54\%}        & 0\%                                                    \\ \hline
\textbf{Phi4-4B}                        & \multicolumn{1}{c|}{800}               & \multicolumn{1}{c|}{554}         & \multicolumn{1}{c|}{\textbf{31\%}$\textcolor{ForestGreen}{\downarrow}$} & \multicolumn{1}{c|}{182}               & \multicolumn{1}{c|}{121}         & \multicolumn{1}{c|}{\textbf{34\%}$\textcolor{ForestGreen}{\downarrow}$} & \multicolumn{1}{c|}{76\%}              & \multicolumn{1}{c|}{76\%}        & 0\%                                      & \multicolumn{1}{c|}{583}               & \multicolumn{1}{c|}{276}         & \multicolumn{1}{c|}{\textbf{53\%}$\textcolor{ForestGreen}{\downarrow}$} & \multicolumn{1}{c|}{134}               & \multicolumn{1}{c|}{61}          & \multicolumn{1}{c|}{\textbf{55\%}$\textcolor{ForestGreen}{\downarrow}$}                 & \multicolumn{1}{c|}{57\%}              & \multicolumn{1}{l|}{57\%}        & 0\%                                                    \\ \hline
\textbf{Qwen3-4B}                       & \multicolumn{1}{c|}{4612}              & \multicolumn{1}{c|}{3255}        & \multicolumn{1}{c|}{\textbf{29\%}$\textcolor{ForestGreen}{\downarrow}$} & \multicolumn{1}{c|}{884}               & \multicolumn{1}{c|}{615}         & \multicolumn{1}{c|}{\textbf{30\%}$\textcolor{ForestGreen}{\downarrow}$} & \multicolumn{1}{c|}{70\%}              & \multicolumn{1}{c|}{76\%}        & 6\%$\textcolor{ForestGreen}{\uparrow}$ & \multicolumn{1}{c|}{4956}              & \multicolumn{1}{c|}{4041}        & \multicolumn{1}{c|}{\textbf{19\%}$\textcolor{ForestGreen}{\downarrow}$} & \multicolumn{1}{c|}{962}               & \multicolumn{1}{c|}{772}         & \multicolumn{1}{c|}{\textbf{20\%}$\textcolor{ForestGreen}{\downarrow}$}                 & \multicolumn{1}{c|}{53\%}              & \multicolumn{1}{l|}{50\%}        & 3\%$\textcolor{red}{\downarrow}$       \\ \hline
\textbf{Qwen2.5-Coder-3B}               & \multicolumn{1}{c|}{1961}              & \multicolumn{1}{c|}{965}         & \multicolumn{1}{c|}{\textbf{51\%}$\textcolor{ForestGreen}{\downarrow}$} & \multicolumn{1}{c|}{493}               & \multicolumn{1}{c|}{210}         & \multicolumn{1}{c|}{\textbf{57\%}$\textcolor{ForestGreen}{\downarrow}$} & \multicolumn{1}{c|}{83\%}              & \multicolumn{1}{c|}{81\%}        & 2\%$\textcolor{red}{\downarrow}$         & \multicolumn{1}{c|}{2696}              & \multicolumn{1}{c|}{955}         & \multicolumn{1}{c|}{\textbf{65\%}$\textcolor{ForestGreen}{\downarrow}$} & \multicolumn{1}{c|}{666}               & \multicolumn{1}{c|}{232}         & \multicolumn{1}{c|}{\textbf{65\%}$\textcolor{ForestGreen}{\downarrow}$}                 & \multicolumn{1}{c|}{56\%}              & \multicolumn{1}{l|}{57\%}        & 1\%$\textcolor{ForestGreen}{\uparrow}$ \\ \hline
\end{tabular}

\caption{Python benchmarks.}
\label{tab:python}
\end{subtable}

\vspace{0.3cm}

\begin{subtable}{\textwidth}
\centering
\begin{tabular}{|l|ccccccclc|ccccccccc|}
\hline
\multicolumn{1}{|c|}{}                   & \multicolumn{9}{c|}{\textbf{HumanEval Java}}                                                                                                                                                                                                                                                                                                                                                                 & \multicolumn{9}{c|}{\textbf{APPS}}                                                                                                                                                                                                                                                                                                                                                                                          \\ \cline{2-19} 
\multicolumn{1}{|c|}{}                   & \multicolumn{3}{c|}{\textbf{GPU Energy}}                                                                                                       & \multicolumn{3}{c|}{\textbf{Output}}                                                                                                       & \multicolumn{3}{c|}{\textbf{Accuracy}}                                                                             & \multicolumn{3}{c|}{\textbf{GPU Energy}}                                                                                                                       & \multicolumn{3}{c|}{\textbf{Output}}                                                                                                      & \multicolumn{3}{l|}{\textbf{Accuracy}}                                                                             \\ \cline{2-19} 
\multicolumn{1}{|c|}{\multirow{-3}{*}{}} & \multicolumn{1}{c|}{\textbf{Baseline}} & \multicolumn{1}{c|}{\textbf{BS}} & \multicolumn{1}{c|}{\textbf{$\Delta$}}                         & \multicolumn{1}{c|}{\textbf{Baseline}} & \multicolumn{1}{c|}{\textbf{BS}} & \multicolumn{1}{c|}{\textbf{$\Delta$}}                         & \multicolumn{1}{c|}{\textbf{Baseline}} & \multicolumn{1}{c|}{\textbf{BS}} & \textbf{$\Delta$}                      & \multicolumn{1}{c|}{\textbf{Baseline}} & \multicolumn{1}{c|}{\textbf{BS}}                  & \multicolumn{1}{c|}{\textbf{$\Delta$}}                        & \multicolumn{1}{c|}{\textbf{Baseline}} & \multicolumn{1}{c|}{\textbf{BS}} & \multicolumn{1}{c|}{\textbf{$\Delta$}}                        & \multicolumn{1}{l|}{\textbf{Baseline}} & \multicolumn{1}{l|}{\textbf{BS}} & \multicolumn{1}{l|}{\textbf{$\Delta$}} \\ \hline
\textbf{CodeLlama-7B}                    & \multicolumn{1}{c|}{1987}              & \multicolumn{1}{c|}{1400}        & \multicolumn{1}{c|}{30\%$\textcolor{ForestGreen}{\downarrow}$} & \multicolumn{1}{c|}{367}               & \multicolumn{1}{c|}{248}         & \multicolumn{1}{c|}{\textbf{32\%}$\textcolor{ForestGreen}{\downarrow}$} & \multicolumn{1}{c|}{31\%}              & \multicolumn{1}{l|}{31\%}        & 0\%                                    & \multicolumn{1}{c|}{1900}              & \multicolumn{1}{c|}{1980} & \multicolumn{1}{c|}{4\%$\textcolor{red}{\uparrow}$}           & \multicolumn{1}{c|}{339}               & \multicolumn{1}{c|}{339}         & \multicolumn{1}{c|}{0\%}                                      & \multicolumn{1}{c|}{1\%}               & \multicolumn{1}{c|}{1\%}         & 0\%                                    \\ \hline
\textbf{Qwen2.5-Coder-7B}                & \multicolumn{1}{c|}{2510}              & \multicolumn{1}{c|}{1387}        & \multicolumn{1}{c|}{\textbf{45\%}$\textcolor{ForestGreen}{\downarrow}$} & \multicolumn{1}{c|}{476}               & \multicolumn{1}{c|}{231}         & \multicolumn{1}{c|}{\textbf{52\%}$\textcolor{ForestGreen}{\downarrow}$} & \multicolumn{1}{c|}{56\%}              & \multicolumn{1}{l|}{56\%}        & 0\%                                    & \multicolumn{1}{c|}{960}               & \multicolumn{1}{c|}{1056}                         & \multicolumn{1}{c|}{10\%$\textcolor{red}{\uparrow}$}          & \multicolumn{1}{c|}{181}               & \multicolumn{1}{c|}{181}         & \multicolumn{1}{c|}{0\%}                                      & \multicolumn{1}{c|}{13\%}              & \multicolumn{1}{c|}{13\%}        & 0\%                                    \\ \hline
\textbf{Deepseek-6.7B}                   & \multicolumn{1}{c|}{5434}              & \multicolumn{1}{c|}{2774}        & \multicolumn{1}{c|}{\textbf{50\%}$\textcolor{ForestGreen}{\downarrow}$} & \multicolumn{1}{c|}{988}               & \multicolumn{1}{c|}{489}         & \multicolumn{1}{c|}{\textbf{51\%}$\textcolor{ForestGreen}{\downarrow}$} & \multicolumn{1}{c|}{55\%}              & \multicolumn{1}{l|}{55\%}        & 0\%                                    & \multicolumn{1}{c|}{5239}              & \multicolumn{1}{c|}{5252} & \multicolumn{1}{c|}{\textbf{1\%}$\textcolor{red}{\uparrow}$}           & \multicolumn{1}{c|}{930}               & \multicolumn{1}{c|}{872}         & \multicolumn{1}{c|}{6\%$\textcolor{ForestGreen}{\downarrow}$} & \multicolumn{1}{c|}{9\%}               & \multicolumn{1}{c|}{9\%}         & 0\%                                    \\ \hline
\textbf{CodeGemma-7B}                    & \multicolumn{1}{c|}{2065}              & \multicolumn{1}{c|}{1573}        & \multicolumn{1}{c|}{24\%$\textcolor{ForestGreen}{\downarrow}$} & \multicolumn{1}{c|}{310}               & \multicolumn{1}{c|}{223}         & \multicolumn{1}{c|}{28\%$\textcolor{ForestGreen}{\downarrow}$} & \multicolumn{1}{c|}{34\%}              & \multicolumn{1}{l|}{34\%}        & 0\%                                    & \multicolumn{1}{c|}{1660}              & \multicolumn{1}{c|}{1749}                         & \multicolumn{1}{c|}{5\%$\textcolor{red}{\uparrow}$}           & \multicolumn{1}{c|}{246}               & \multicolumn{1}{c|}{246}         & \multicolumn{1}{c|}{0\%}                                      & \multicolumn{1}{c|}{6\%}               & \multicolumn{1}{c|}{6\%}         & 0\%                                    \\ \hline
\textbf{CodeQwen1.5-7B}                  & \multicolumn{1}{c|}{2995}              & \multicolumn{1}{c|}{1328}        & \multicolumn{1}{c|}{\textbf{56\%}$\textcolor{ForestGreen}{\downarrow}$} & \multicolumn{1}{c|}{573}               & \multicolumn{1}{c|}{244}         & \multicolumn{1}{c|}{\textbf{57\%}$\textcolor{ForestGreen}{\downarrow}$} & \multicolumn{1}{c|}{48\%}              & \multicolumn{1}{l|}{48\%}        & 0\%                                    & \multicolumn{1}{c|}{1212}              & \multicolumn{1}{c|}{1279}                         & \multicolumn{1}{c|}{6\%$\textcolor{red}{\uparrow}$}           & \multicolumn{1}{c|}{230}               & \multicolumn{1}{c|}{230}         & \multicolumn{1}{c|}{0\%}                                      & \multicolumn{1}{c|}{10\%}              & \multicolumn{1}{c|}{10\%}        & 0\%                                    \\ \hline
\textbf{NextCoder-7B}                    & \multicolumn{1}{c|}{486}               & \multicolumn{1}{c|}{388}         & \multicolumn{1}{c|}{20\%$\textcolor{ForestGreen}{\downarrow}$} & \multicolumn{1}{c|}{93}                & \multicolumn{1}{c|}{72}          & \multicolumn{1}{c|}{23\%$\textcolor{ForestGreen}{\downarrow}$} & \multicolumn{1}{c|}{71\%}              & \multicolumn{1}{l|}{71\%}        & 0\%                                    & \multicolumn{1}{c|}{1347}              & \multicolumn{1}{c|}{1377}                         & \multicolumn{1}{c|}{2\%$\textcolor{red}{\uparrow}$}           & \multicolumn{1}{c|}{255}               & \multicolumn{1}{c|}{240}         & \multicolumn{1}{c|}{6\%$\textcolor{ForestGreen}{\downarrow}$} & \multicolumn{1}{c|}{20\%}              & \multicolumn{1}{c|}{20\%}        & 0\%                                    \\ \hline
\textbf{Phi3.5-4B}                       & \multicolumn{1}{c|}{1651}              & \multicolumn{1}{c|}{1073}        & \multicolumn{1}{c|}{\textbf{35\%}$\textcolor{ForestGreen}{\downarrow}$} & \multicolumn{1}{c|}{394}               & \multicolumn{1}{c|}{229}         & \multicolumn{1}{c|}{\textbf{42\%}$\textcolor{ForestGreen}{\downarrow}$} & \multicolumn{1}{c|}{39\%}              & \multicolumn{1}{l|}{39\%}        & 0\%                                    & \multicolumn{1}{c|}{1867}              & \multicolumn{1}{c|}{2035}                         & \multicolumn{1}{c|}{9\%$\textcolor{red}{\uparrow}$}           & \multicolumn{1}{c|}{447}               & \multicolumn{1}{c|}{436}         & \multicolumn{1}{c|}{3\%$\textcolor{ForestGreen}{\downarrow}$} & \multicolumn{1}{c|}{6\%}               & \multicolumn{1}{c|}{6\%}         & 0\%                                    \\ \hline
\textbf{Phi4-4B}                         & \multicolumn{1}{c|}{3939}              & \multicolumn{1}{c|}{2631}        & \multicolumn{1}{c|}{33\%$\textcolor{ForestGreen}{\downarrow}$} & \multicolumn{1}{c|}{901}               & \multicolumn{1}{c|}{517}         & \multicolumn{1}{c|}{\textbf{43\%}$\textcolor{ForestGreen}{\downarrow}$} & \multicolumn{1}{c|}{46\%}              & \multicolumn{1}{l|}{46\%}        & 0\%                                    & \multicolumn{1}{c|}{1398}              & \multicolumn{1}{c|}{1530}                         & \multicolumn{1}{c|}{9\%$\textcolor{red}{\uparrow}$}           & \multicolumn{1}{c|}{320}               & \multicolumn{1}{c|}{320}         & \multicolumn{1}{c|}{0\%}                                      & \multicolumn{1}{c|}{10\%}              & \multicolumn{1}{c|}{10\%}        & 0\%                                    \\ \hline
\textbf{Qwen3-4B}                        & \multicolumn{1}{c|}{5271}              & \multicolumn{1}{c|}{2725}        & \multicolumn{1}{c|}{\textbf{48\%}$\textcolor{ForestGreen}{\downarrow}$} & \multicolumn{1}{c|}{1000}              & \multicolumn{1}{c|}{466}         & \multicolumn{1}{c|}{\textbf{53\%}$\textcolor{ForestGreen}{\downarrow}$} & \multicolumn{1}{c|}{60\%}              & \multicolumn{1}{l|}{62\%}        & 2\%$\textcolor{ForestGreen}{\uparrow}$ & \multicolumn{1}{c|}{4964}              & \multicolumn{1}{c|}{4940}                         & \multicolumn{1}{c|}{1\%$\textcolor{ForestGreen}{\downarrow}$} & \multicolumn{1}{c|}{939}               & \multicolumn{1}{c|}{900}         & \multicolumn{1}{c|}{4\%$\textcolor{ForestGreen}{\downarrow}$} & \multicolumn{1}{c|}{3\%}               & \multicolumn{1}{l|}{4\%}         & 1\%$\textcolor{ForestGreen}{\uparrow}$ \\ \hline
\textbf{Qwen2.5-Coder-3B}                & \multicolumn{1}{c|}{2626}              & \multicolumn{1}{c|}{1008}        & \multicolumn{1}{c|}{\textbf{62\%}$\textcolor{ForestGreen}{\downarrow}$} & \multicolumn{1}{c|}{608}               & \multicolumn{1}{c|}{242}         & \multicolumn{1}{c|}{\textbf{60\%}$\textcolor{ForestGreen}{\downarrow}$} & \multicolumn{1}{c|}{69\%}              & \multicolumn{1}{l|}{70\%}        & 1\%$\textcolor{ForestGreen}{\uparrow}$ & \multicolumn{1}{c|}{3746}               & \multicolumn{1}{c|}{3654}                          & \multicolumn{1}{c|}{\textbf{7\%}$\textcolor{ForestGreen}{\downarrow}$} & \multicolumn{1}{c|}{917}              & \multicolumn{1}{c|}{849}        & \multicolumn{1}{c|}{2\%$\textcolor{ForestGreen}{\downarrow}$} & \multicolumn{1}{c|}{5\%}               & \multicolumn{1}{c|}{7\%}         & 2\%$\textcolor{ForestGreen}{\uparrow}$ \\ \hline
\end{tabular}
\caption{Java benchmarks.}
\label{tab:java}
\end{subtable}

\caption{\small
Mean total GPU energy consumption, output length, and accuracy for baseline code generation and code generation with \textit{Babbling Suppression} (BS). GPU Energy is reported in Joules, output length is measured in number of tokens, and accuracy is pass@1. Mean values are calculated separately for each Java and Python benchmarks. Each metric also includes a $\Delta$ value, representing the difference between the baseline and BS. \textcolor{red}{Red} and \textcolor{ForestGreen}{green}
colors denote negative and positive impacts of \textit{BS}, respectively. Values in \textbf{boldface} indicate a statistically-significant difference, p-value < 0.005, using Mann-Whitney U test for GPU energy and output and Two-proportion Z-test for accuracy.}
\label{tab:EnergyAccuracy}
\end{table*}
}

\newcommand{\tableEpT}{
\begin{table*}[]
\centering
\footnotesize
\begin{tabular}{|l|ccc|ccc|clc|ccc|}
\hline
\multicolumn{1}{|c|}{}                   & \multicolumn{3}{c|}{\textbf{HumanEval}}                                                                     & \multicolumn{3}{c|}{\textbf{MBPP}}                                                                          & \multicolumn{3}{c|}{\textbf{HumanEval Java}}                                                                          & \multicolumn{3}{c|}{\textbf{APPS}}                                                                                                    \\ \cline{2-13} 
\multicolumn{1}{|c|}{\multirow{-2}{*}{}} & \multicolumn{1}{c|}{\textbf{Baseline}} & \multicolumn{1}{c|}{\textbf{BS}} & \textbf{$\Delta$}               & \multicolumn{1}{c|}{\textbf{Baseline}} & \multicolumn{1}{c|}{\textbf{BS}} & \textbf{$\Delta$}               & \multicolumn{1}{c|}{\textbf{Baseline}} & \multicolumn{1}{c|}{\textbf{BS}} & \textbf{$\Delta$}                         & \multicolumn{1}{c|}{\textbf{Baseline}} & \multicolumn{1}{c|}{\textbf{BS}}                  & \textbf{$\Delta$}                        \\ \hline
\textbf{CodeLlama-7B}                    & \multicolumn{1}{c|}{5.32}              & \multicolumn{1}{c|}{5.61}        & \textbf{6\%}$\textcolor{red}{\uparrow}$  & \multicolumn{1}{c|}{5.37}              & \multicolumn{1}{c|}{5.68}        & \textbf{6\%}$\textcolor{red}{\uparrow}$  & \multicolumn{1}{c|}{5.34}              & \multicolumn{1}{l|}{5.57}        & 4\%$\textcolor{red}{\uparrow}$            & \multicolumn{1}{c|}{5.55}              & \multicolumn{1}{c|}{6.29} & \textbf{13\%}$\textcolor{red}{\uparrow}$          \\ \hline
\textbf{Qwen2.5-Coder-7B}                & \multicolumn{1}{c|}{5.19}              & \multicolumn{1}{c|}{5.54}        & \textbf{7\%}$\textcolor{red}{\uparrow}$  & \multicolumn{1}{c|}{5.07}              & \multicolumn{1}{c|}{5.66}        & \textbf{12\%}$\textcolor{red}{\uparrow}$ & \multicolumn{1}{c|}{5.27}              & \multicolumn{1}{l|}{6.06}        & 15\%$\textcolor{red}{\uparrow}$           & \multicolumn{1}{c|}{5.33}              & \multicolumn{1}{c|}{6.02}                         & \textbf{13\%}$\textcolor{red}{\uparrow}$          \\ \hline
\textbf{Deepseek-6.7B}                   & \multicolumn{1}{c|}{5.44}              & \multicolumn{1}{c|}{5.83}        & 7\%$\textcolor{red}{\uparrow}$  & \multicolumn{1}{c|}{5.36}              & \multicolumn{1}{c|}{5.61}        & \textbf{5\%}$\textcolor{red}{\uparrow}$  & \multicolumn{1}{c|}{5.51}              & \multicolumn{1}{l|}{5.31}        & 4\%$\textcolor{ForestGreen}{\downarrow}$  & \multicolumn{1}{c|}{5.62}              & \multicolumn{1}{c|}{5.98} & \textbf{6\%}$\textcolor{red}{\uparrow}$           \\ \hline
\textbf{CodeGemma-7B}                    & \multicolumn{1}{c|}{6.43}              & \multicolumn{1}{c|}{7.37}        & \textbf{15\%}$\textcolor{red}{\uparrow}$ & \multicolumn{1}{c|}{6.71}              & \multicolumn{1}{c|}{7.51}        & \textbf{12\%}$\textcolor{red}{\uparrow}$ & \multicolumn{1}{c|}{6.61}              & \multicolumn{1}{l|}{8.22}        & \textbf{24\%}$\textcolor{red}{\uparrow}$           & \multicolumn{1}{c|}{6.73}              & \multicolumn{1}{c|}{7.32}                         & \textbf{9\%}$\textcolor{red}{\uparrow}$           \\ \hline
\textbf{CodeQwen1.5-7B}                  & \multicolumn{1}{c|}{5.19}              & \multicolumn{1}{c|}{5.57}        & \textbf{7\%}$\textcolor{red}{\uparrow}$  & \multicolumn{1}{c|}{5.12}              & \multicolumn{1}{c|}{5.46}        & \textbf{7\%}$\textcolor{red}{\uparrow}$  & \multicolumn{1}{c|}{5.23}              & \multicolumn{1}{l|}{5.32}        & 2\%$\textcolor{red}{\uparrow}$            & \multicolumn{1}{c|}{5.28}              & \multicolumn{1}{c|}{5.72}                         & \textbf{8\%}$\textcolor{red}{\uparrow}$           \\ \hline
\textbf{NextCoder-7B}                    & \multicolumn{1}{c|}{5.22}              & \multicolumn{1}{c|}{5.46}        & \textbf{5\%}$\textcolor{red}{\uparrow}$  & \multicolumn{1}{c|}{5.22}              & \multicolumn{1}{c|}{5.61}        & \textbf{8\%}$\textcolor{red}{\uparrow}$  & \multicolumn{1}{c|}{5.22}              & \multicolumn{1}{l|}{5.43}        & \textbf{4\%}$\textcolor{red}{\uparrow}$            & \multicolumn{1}{c|}{5.33}              & \multicolumn{1}{c|}{5.87}                         & \textbf{10\%}$\textcolor{red}{\uparrow}$          \\ \hline
\textbf{Phi3.5-4B}                       & \multicolumn{1}{c|}{4.13}              & \multicolumn{1}{c|}{4.46}        & \textbf{8\%}$\textcolor{red}{\uparrow}$  & \multicolumn{1}{c|}{4.29}              & \multicolumn{1}{c|}{4.69}        & 9\%$\textcolor{red}{\uparrow}$  & \multicolumn{1}{c|}{4.16}              & \multicolumn{1}{l|}{4.76}        & 14\%$\textcolor{red}{\uparrow}$           & \multicolumn{1}{c|}{4.18}              & \multicolumn{1}{c|}{4.91}                         & \textbf{18\%}$\textcolor{red}{\uparrow}$          \\ \hline
\textbf{Phi4-4B}                         & \multicolumn{1}{c|}{4.32}              & \multicolumn{1}{c|}{4.61}        & \textbf{7\%}$\textcolor{red}{\uparrow}$  & \multicolumn{1}{c|}{4.36}              & \multicolumn{1}{c|}{4.44}        & \textbf{2\%}$\textcolor{red}{\uparrow}$  & \multicolumn{1}{c|}{4.39}              & \multicolumn{1}{l|}{4.61}        & \textbf{5\%}$\textcolor{red}{\uparrow}$            & \multicolumn{1}{c|}{4.35}              & \multicolumn{1}{c|}{5.12}                         & \textbf{18\%}$\textcolor{red}{\uparrow}$          \\ \hline
\textbf{Qwen3-4B}                        & \multicolumn{1}{c|}{5.21}              & \multicolumn{1}{c|}{5.37}        & 3\%$\textcolor{red}{\uparrow}$  & \multicolumn{1}{c|}{5.15}              & \multicolumn{1}{c|}{5.35}        & \textbf{4\%}$\textcolor{red}{\uparrow}$  & \multicolumn{1}{c|}{5.27}              & \multicolumn{1}{l|}{5.59}        & \textbf{6\%}$\textcolor{red}{\uparrow}$            & \multicolumn{1}{c|}{5.28}              & \multicolumn{1}{c|}{5.49}                         & \textbf{4\%}$\textcolor{red}{\uparrow}$ \\ \hline
\textbf{Qwen2.5-Coder-3B}                & \multicolumn{1}{c|}{3.97}              & \multicolumn{1}{c|}{4.71}        & \textbf{19\%}$\textcolor{red}{\uparrow}$ & \multicolumn{1}{c|}{4.05}              & \multicolumn{1}{c|}{4.21}        & \textbf{4\%}$\textcolor{red}{\uparrow}$  & \multicolumn{1}{c|}{4.29}              & \multicolumn{1}{l|}{3.86}        & \textbf{10\%}$\textcolor{ForestGreen}{\downarrow}$ & \multicolumn{1}{c|}{4.09}              & \multicolumn{1}{c|}{4.31}                         & \textbf{5\%}$\textcolor{red}{\uparrow}$ \\ \hline
\end{tabular}
\caption{\small
Mean GPU energy per token for baseline code generation and code generation with \textit{babbling suppression (BS)}, reported in Joules. Values represent the average across Python and Java benchmarks. $\Delta$ indicates the difference between the baseline and BS, with \textcolor{red}{red} and \textcolor{ForestGreen}{green} highlighting negative and positive effects of BS, respectively. Values in \textbf{boldface} indicate a statistically-significant difference, p-value < 0.005, using Mann-Whitney U test.}
\label{tab:eptAll}
\end{table*}
}

\newcommand{\tableOverhead}{
\begin{table}[t!]
\centering
\scriptsize
\begin{tabular}{|l|cc|cc|cc|}
\hline
\multirow{2}{*}{} & \multicolumn{2}{c|}{\textbf{CodeLlama-7B}} & \multicolumn{2}{c|}{\textbf{Qwen2.5-Coder-7B}} & \multicolumn{2}{c|}{\textbf{Phi4-4B}} \\ \cline{2-7} 
 & \textbf{Baseline} & \textbf{BS} & \textbf{Baseline} & \textbf{BS} & \textbf{Baseline} & \textbf{BS} \\ \hline
\textbf{Output} & 600 & 461 & 120 & 118 & 182 & 121 \\ \hline
\textbf{EpT (J)} & 5.32 & 5.61 & 5.19 & 5.54 & 4.32 & 4.61 \\ \hline
\textbf{Time to token (s)} & 0.036 & 0.041 & 0.035 & 0.041 & 0.031 & 0.037 \\ \hline
\textbf{BS time (s)} & -- & 0.1266 & -- & 0.1961 & -- & 0.1755 \\ \hline
\textbf{Total time (s)} & 21.79 & 17.76 & 4.26 & 5.17 & 5.32 & 4.55 \\ \hline
\textbf{GPU Energy (J)} & 3242 & 2578 & 625 & 672 & 800 & 554 \\ \hline
\textbf{GPU Power (W)} & 149 & 144 & 148 & 144 & 141 & 132 \\ \hline
\textbf{Avg. \# of BS tests } & -- & 2.96 & -- & 2.53 & -- & 2.68 \\ \hline
\end{tabular}
\caption{
\small
Results for overhead analysis: output is measured in number of tokens; EpT denotes energy per token (Joules); time to token represents the time required to generate a token; BS time is the duration of the algorithm; total time corresponds to a single inference run; Avg. \#  of BS tests indicates the average number of test executions by BS per inference. The results are averaged over the HumanEval benchmark. }
\label{tab:overhead}
\end{table}
}
\newcommand{\bsalgorithm}{
\begin{algorithm}[t]
\caption{General steps of Babbling Suppression}
\label{alg:bs}
\begin{algorithmic}[1]
\Require test\_cases, decoded LLM output
\State $discarded\_check\_units = []$
\State $generated\_sequence =$ ``''
\While{LLM generates tokens}
    \State $generated\_sequence$ += $next$ $generated$ $token$
\State $checking\_units = $\par \hskip\algorithmicindent$form\_checking\_units(generated\_sequence)$
    \State $functions = $``''

    \ForAll{$unit \in checking\_units$}
        \If{$unit \notin discarded\_check\_units$}
            \If{$check\_well\_formedness(functions+unit)$}
                \State $functions$ += $unit$
                \If{$run\_tests(functions)$}
                    \State $stop$ $generation$
                \EndIf
            \Else 
                \If{$should\_discard(unit)$}
                    \State $discarded\_check\_units.add(unit)$
                \Else 
                     \State $functions$ += $unit$
                \EndIf
            \EndIf
        \EndIf
    \EndFor
\EndWhile
\end{algorithmic}
\end{algorithm}
}

\usepackage{multirow}
 
\setcopyright{acmlicensed}
\copyrightyear{2026}
\acmYear{2026}

\usepackage[breakable, skins]{tcolorbox}
\usepackage[dvipsnames]{xcolor}
\usepackage{soul}
\usepackage{algorithm}
\usepackage{algpseudocode}
\usepackage{float}
\tcbuselibrary{skins}
\usepackage{listings}
\usepackage{xcolor}
\usepackage{multirow}
\usepackage{subcaption}
\usepackage{textcomp}
\newenvironment{summarybox}
{\begin{tcolorbox}
[breakable,enhanced,arc=0mm,colback=gray!10,frame hidden,overlay broken={%
    \draw[thick,black] (interior.north west)--(interior.south west);
},overlay unbroken={%
    \draw[thick,black] (interior.north west)--(interior.south west);
},left=2pt,right=0pt,top=0pt,bottom=0pt,before={\vspace{3pt}\noindent},after={\vspace{0pt}}]
\setlength{\baselineskip}{0.75\baselineskip}}
{\end{tcolorbox}}
\newenvironment{summary}
{\vspace{5pt}\noindent\begin{summarybox}}
{\end{summarybox}\vspace{-5pt}}

\begin{document}
\renewcommand{\hl}[1]{#1}
\title{Babbling Suppression: Making LLMs Greener One Token at a Time}
%

\author{Lola Solovyeva}
\email{o.solovyeva@utwente.nl}
\orcid{0009-0008-6903-7086}
\affiliation{%
  \institution{University of Twente}
  \city{Enschede}
  \country{the Netherlands}
}

\author{Fernando Castor}
\email{f.castor@utwente.nl}
\affiliation{%
  \institution{University of Twente}
  \city{Enschede}
  \country{the Netherlands}
}


\begin{abstract}
\textbf{Context:} Large Language Models (LLMs) are increasingly used in modern software development, aiding in code generation, code completion, and refactoring through AI-powered assistants. While they accelerate development workflows, they often produce extraneous output, referred to as \textbf{"babbling"}, which incurs additional cognitive, economic, and energy costs.
\textbf{Objective:} This work investigates the problem of babbling in LLM-based code generation and proposes a practical, model-agnostic approach to reduce unnecessary output without compromising solution accuracy. 
\textbf{Method:} We introduce Babbling Suppression (BS), a method that integrates test execution into the LLM generation process by evaluating intermediate outputs and terminating generation once a solution passes all tests. This prevents excessive token generation while having no impact on model accuracy. An empirical study was conducted across two Python and two Java benchmarks, targeting four 3–4B parameter models and six 6–7B parameter models.
\textbf{Results:} Our findings show that babbling occurs across all tested models, with higher frequency in Java than in Python. Applying BS significantly reduces  energy consumption by up to 65\% for Python and 62\% for Java in models prone to babbling. Across 40 model-benchmark pairs, 29 showed reduced mean energy consumption, with reductions exceeding 20\% in 22 cases. Generated token count decreased in 35 pairs, while the GPU energy-per-token overhead of BS remained below 10\% for 26 pairs, decreased for 2, and reached a maximum of 24\%, yielding net energy savings in most cases.
\textbf{Implications:} BS can make AI-assisted programming more efficient and sustainable by reducing energy consumption and minimizing cognitive effort by developers. Its model-agnostic design allows easy integration, suggesting broad applicability. The replication package is available on Zenodo~\cite{replicationPackage}.

\end{abstract}

\maketitle

\section{Introduction}
\par Large Language Models (LLMs) have rapidly become  part of modern software development workflows, particularly for code generation tasks~\cite{SERGEYUK2025107610, 10403378, 10.1145/3747588, REVURI2026115}. AI-powered assistants such as GitHub Copilot~\cite{githubCopilot} and Claude~\cite{claude} aid developers in generating, completing~\cite{HUSEIN2025103917}, and refactoring code~\cite{10.1145/3801158}, demonstrating their potential to accelerate development processes. These systems are being embedded into development environments, transforming how developers interact with code and reshaping parts of the programming process into a collaboration between humans and AI~\cite{10.1145/3643795.3648379}.

The use of LLMs in software development, in particular to generate code, comes with caveats. A key concern is the inaccuracy of generated outputs and the resulting lack of trust. This has received significant attention in the literature~\cite{humanEval,KUHAIL2024103111,SERGEYUK2025107610,10.1145/3613904.3641936,10.1145/3747588}. Even when LLMs produce correct solutions, they often output additional tokens with explanations, tests, prototype solutions, code comments, and more. These extra tokens are typically removed during post-processing or simply ignored by the developer. We call this behavior \textbf{``babbling''}. It has received  little attention in the literature. 

\par The problem of babbling is relevant, as it brings additional costs to the use of LLMs. First, because LLMs are energy-intensive. Small, locally-deployable language models can deplete a contemporary laptop's battery in a few hundred inferences~\cite{alizadeh2025languagemodelssoftwaredevelopment}. Frontier LLMs made available as services, e.g., ChatGPT, consume so much energy that their service fees are not sufficient to cover their operational costs~\cite{Economist:2025:OLM,Economist:2025:WOE}. In both cases, the energy use of the model grows proportionally to the number output tokens it produces. Second, LLM services are usually billed based on the number of processed (input, output) tokens. If an LLM response produces more tokens than necessary, this costs more money for the customers. Third, because of cognitive effort~\cite{Tang:2024:DBV,Zhang:2024:VVD}. When a developer asks an LLM for a solution to a programming problem, having to weed out extra tokens is taxing and consumes time that could be used more productively. Reducing babbling is therefore beneficial, as it impacts these three costs, energetic, economic, and cognitive, as long as accuracy is not compromised. Previous work has attempted to address babbling. For example, Guo et al.~\cite{10.1145/3650212.3680343} employ a lightweight module attached to the LLM to detect when excessive tokens are being generated. This  requires retraining the module for each new model, as well as for different tasks and programming languages.

We propose an approach named Babbling Suppression (BS) to reduce the costs of using LLMs to generate code. BS integrates test execution into the LLM generation process by evaluating intermediate outputs and terminating generation once a solution passes all tests. This approach stems from four basic observations: (i) babbling primarily occurs \textit{after} a solution has been produced, (ii) for code, it is possible to objectively verify generated solutions by testing, 
(iii) checking whether solutions are correct by testing is often cheaper than generating them, and (iv) discarding extraneous tokens does not compromise accuracy. BS follows an iterative cycle in which tests are written first, and the generated code must satisfy these tests to be considered correct. When all tests pass, code generation immediately ends. By enforcing predefined requirements through tests, it helps ensure that the resulting code aligns with the intended functionality while reducing the number of generated tokens and energy use. This approach can be implemented as a plugin to the code generation process, is model-agnostic, and does not require model retraining or special prompting. 

We conducted an empirical study to evaluate the usefulness of \textit{babbling suppression} for function-level code generation in Python and Java. The study targeted four coding LLMs in the 3–4B range and six coding LLMs in the 6–7B range, using them to generate code for two Python benchmarks and two Java benchmarks. Our findings show that all models can be prone to babbling. Fewer models babble in Python, whereas it happens more frequently in Java. Applying BS mitigated the generation of excessive tokens without affecting model accuracy, reducing average GPU energy consumption by up to 65\% for Python and 62\% for Java, particularly for models prone to babbling. Considering all the 40 model-benchmark pairs, there was a reduction in mean energy consumption for 29 of them and the reduction was 20\% or higher for 22. The number of generated tokens decreased for 35 of the pairs and remained constant for the remaining 5. The energy-per-token overhead introduced by BS was lower than 10\% for 26 of these pairs, decreased for 2, and peaked at 24\% for the remaining 12. In most cases the reduction in generated tokens outweighs the costs, resulting in overall energy savings. The replication package is available on Zenodo~\cite{replicationPackage}.
\begin{figure*}[t!]
    \centering
    \begin{subfigure}{\linewidth}
        \centering
        \includegraphics[width=\linewidth]{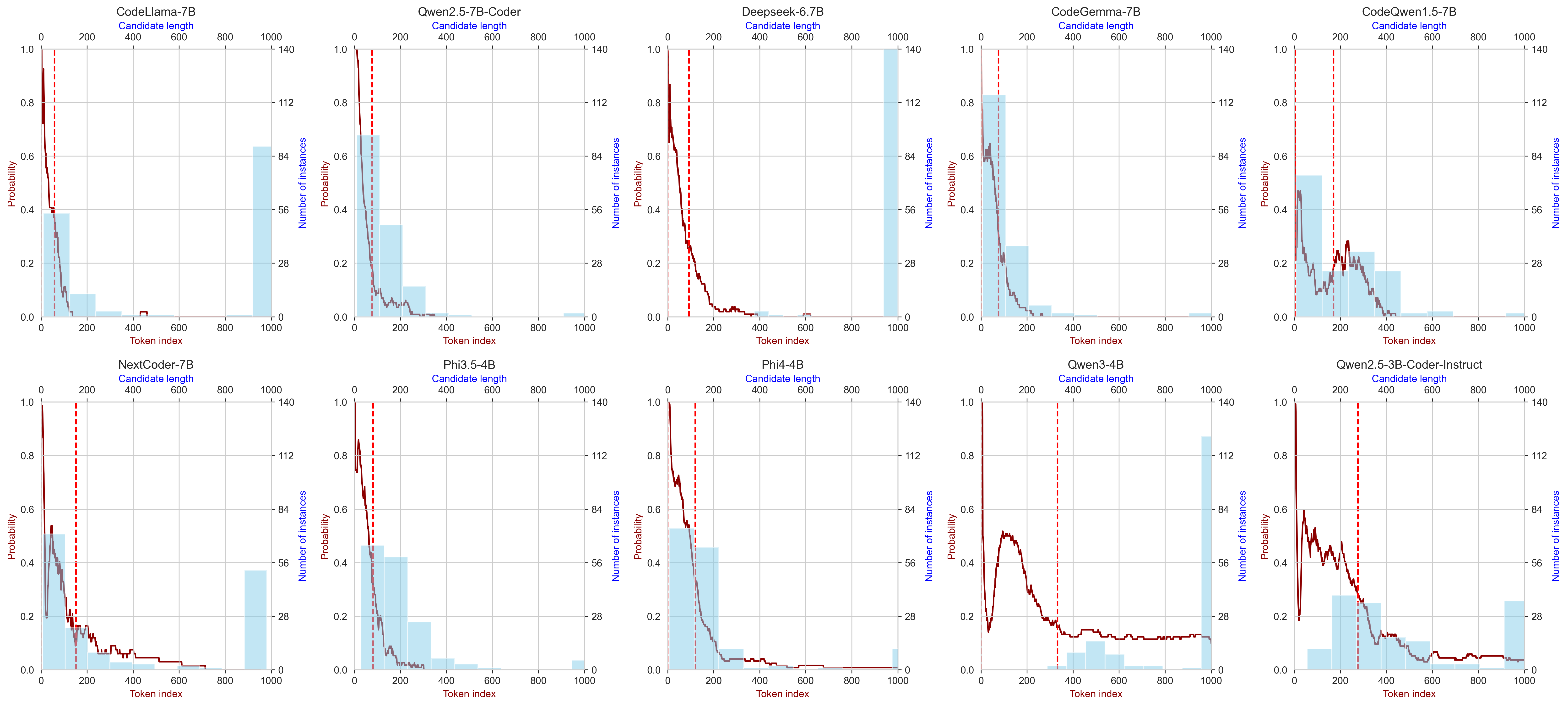}
        \caption{Generated solutions in Python.}
        \label{fig:pythonUsefulTokens}
    \end{subfigure}

    \vspace{0.3cm}

    \begin{subfigure}{\linewidth}
        \centering
        \includegraphics[width=\linewidth]{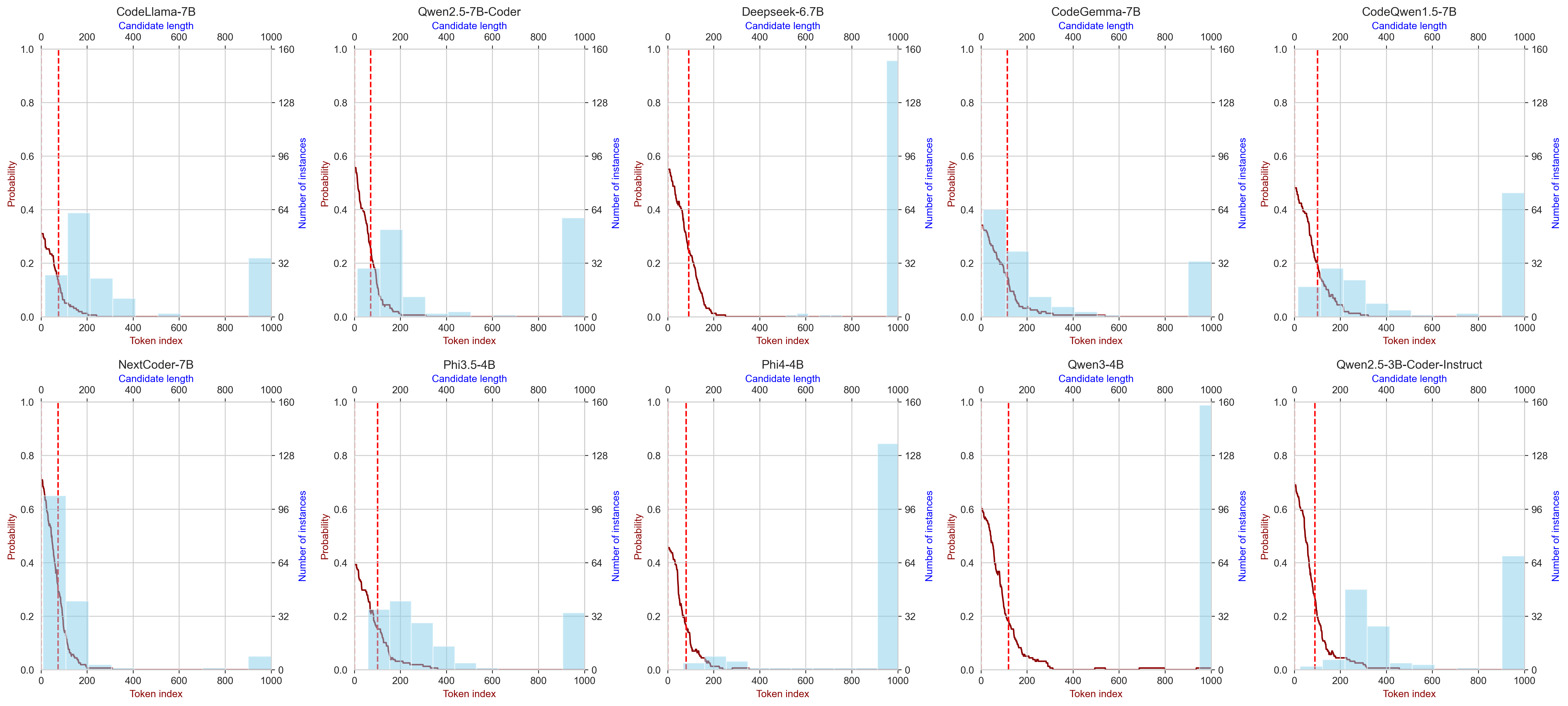}
        \caption{Generated solutions in Java.}
        \label{fig:javaUsefulTokens}
    \end{subfigure}

    \caption{
    \small
    Token likelihood in test-passing solutions, together with histograms showing the lengths of the generated solutions. The bottom x-axis shows the token index, and the left y-axis shows the probability that a token appears in a correctly generated solution. The top x-axis shows the length of the generated solutions, and the right y-axis shows the number of instances with that length.}
    \label{fig:usefulTokens}
\end{figure*}
\section{Related work}
\par The adoption of LLMs in software development has led to a substantial body of research focusing on their effectiveness~\cite{10628428, codeQualityLLM, ai-powered-power-hungry, alizadeh2025languagemodelssoftwaredevelopment}, reliability~\cite{hallucinationsCodeGen, wattsLuccioni}, and impact on developer productivity~\cite{SERGEYUK2025107610, 10.1145/3639477.3643648, GenAISoftware}. In particular, significant attention has been devoted to their application in code generation~\cite{10.1145/3747588, 10.1145/3660810}, with studies examining the correctness~\cite{10.1145/3691620.3695470}, quality~\cite{apsan2025}, and security~\cite{11429547} of generated code across different programming languages.
\par Prior work evaluates LLM-generated code using standardized benchmarks. HumanEval~\cite{humanEval} and MBPP~\cite{austin2021programsynthesislargelanguage} assess functional correctness via unit tests, while APPS~\cite{hendrycks2021measuringcodingchallengecompetence} and LiveBench~\cite{white2025livebenchchallengingcontaminationlimitedllm} cover different levels of difficulty, from basic to competition-level problems. Newer Python benchmarks like COFFE~\cite{10.1145/3691620.3695470} also measure time efficiency using stress tests and CPU instruction-based metrics. Most benchmarks focus on Python due to its popularity and ease of testing. Recent efforts have expanded to other languages: MultiPL-E~\cite{10.1109/TSE.2023.3267446} extends HumanEval to Java, C++, and JavaScript; CodeContests~\cite{wang2025codecontestshighqualitytestcase} offers multilingual problems from competitive programming. JavaBench~\cite{10.1145/3691620.3695470} specifically targets Java, emphasizing project-level generation and object-oriented features, with high test coverage for complex evaluation beyond function-level tasks.
\par These benchmarks typically follow a common paradigm in which a model generates a complete solution that is evaluated only after the generation process has finished. In contrast, our work leverages the same test-based evaluation setting but integrates it directly into the generation loop. Specifically, BS uses the available tests not only as a post-hoc correctness signal, but as an online criterion to determine whether generation can be terminated early. In this sense, our approach is complementary to existing benchmarks. Rather than proposing a new evaluation dataset, we modify how existing test suites are utilized during inference.
\par  Some prior work has explored incorporating tests into LLM-based code generation. Mathews et al.~\cite{10.1145/3691620.3695527} introduced the TGen framework, which operates in a continuous self-improving loop. Developers provide problem statements and test cases, which are then included in the prompt to guide code generation. An LLM-based feedback mechanism iteratively refines the output to ensure all unit tests pass. Similary Piya et al.~\cite{10.1145/3643795.3648382} proposes an incremental development process where users gradually present unit tests to the LLM, guiding it to generate code that passes each test. In our work, we utilize TDD for code generation, but rather than appending tests to the prompt to improve accuracy of the generated code, we use them to truncate unnecessary tokens and reduce the energy consumption of LLM inference.
\par As LLMs become more integrated into software development workflows, it is also important to consider their energy footprint. 
For example, Alizadeh et al.~\cite{alizadeh2025languagemodelssoftwaredevelopment} investigated the performance of 18 families of LLMs on typical software development tasks considering full-precision and quantized versions. Their results show that larger models with higher energy budgets do not always show substantially improved accuracy, and quantized versions of large models can often achieve better efficiency without compromising performance. Similarly, Mehditabar et al.~\cite{mehditabar2025smartcostlybenchmarkingllms} propose the BRACE framework to systematically benchmark code language models on functional correctness and energy efficiency. By evaluating 22 state-of-the-art models, their framework provides insights into the accuracy-energy trade-offs. Our method complements these works by preserving the same level of accuracy while further reducing the energy footprint, providing a practical approach to more sustainable code generation.
\par Lastly, there were previous efforts to perform early exiting strategies to reduce the energy consumption of LLM inference for code generation. GREEN-CODE~\cite{ilager2025greencodelearningoptimizeenergy} trains a reinforcement learning agent to balance the trade-offs between accuracy, latency, and energy consumption. It performs an early exit on the intermediate layers, demonstrating 23–50\% reductions in energy usage on code generation tasks without significantly impacting accuracy. Similarly, CodeFast~\cite{10.1145/3650212.3680343} accelerates code generation by detecting unnecessary tokens. It uses a lightweight model, GenGuard, trained on automatically constructed data to predict when inference can be terminated early. CodeFast improves inference speed across multiple LLMs by 34–452\% while preserving code quality. Another approach focused at reducing unnecessary tokens was proposed by Pan et al.~\cite{pan2025hiddencostreadabilitycode}, who observed that LLMs may not require code formatting to the same extent as humans, as for whom it primarily improves readability. Their study showed that removing non-essential formatting reduces input tokens by an average of 24.5\%, while additional prompting or fine-tuning can decrease output length by up to 36.1\% without affecting correctness, thereby improving LLM efficiency.
\par Our method differs from those works mentioned above in that it does not require additional training, learning agents, or modifications to the input. It performs intermediate checks directly on the output and operates entirely outside the LLM. It avoids the computational and data overhead associated with training auxiliary models and does not require developers to adjust their input. Furthermore, it is model-agnostic and can be applied to any LLM without requiring access to its internal weights or modifying its architecture. Finally, it provides interpretable and deterministic signals for early stopping, ensuring that only valid solutions trigger termination. As a result, energy savings are achieved without compromising accuracy, while maintaining a simpler and more flexible approach.
\section{Motivation}
\label{sec:motivation}
The main motivation for this work stems from the observation that language models tend to generate more content than requested. In the context of function-level code generation, models often produce not only the target function but also additional material, such as explanations, usage examples, alternative implementations, prototypes, test cases, etc. When only the function is required, this extra content is typically discarded during post-processing, but the computational cost of generating it is still incurred. While some of this additional output may occasionally be useful, e.g., to improve understanding, there are also cases where the model generates irrelevant or nonsensical tokens until the maximum output limit is reached. This behavior leads to unnecessary computation without improving the quality of the solution. We refer to this behavior as \textit{babbling}, borrowing the term from the Merriam–Webster dictionary, where it is defined as "the production of meaningless strings of speech"~\cite{babble}. 

\par We conducted a preliminary experiment to verify two hypotheses relevant for this work: (i) models tend to babble beyond producing a correct solution, and (ii) babbling most commonly occurs after the correct solution has been generated. To investigate this, we prompted 10 models for function-level code generation using the HumanEval benchmark for Python and Java. We set the maximum output length to 1000 tokens, which provides sufficient room for generating a correct solution (canonical solutions are typically in the 100–200 tokens range) as well as for observing whether the models attempt to reach the maximum token limit. 
\par Figure~\ref{fig:usefulTokens} presents the results of our preliminary experiments. The plots include two vertical dashed lines: the left line indicates the average position of the first token of the correct solution, while the right line indicates the average position of the last token of the correctly generated solution. These lines show that the number of tokens required to produce a correct solutions tends to be much lower than the 1000-token limit. The dark red line on the plot shows the likelihood of token $n$, where $n$ represents the position of the token in the generated sequence, appearing in a solution that passes all test cases from the benchmark. The bottom x-axis shows the index of the token in the generated sequence, while the left y-axis shows the probability that a token at that position appears in correct solutions. Furthermore, the figure includes blue histograms that show the number of solutions for each output length. The top x-axis shows the output length of the solution, while the right y-axis shows the number of instances with that length. 
\par From Figure~\ref{fig:pythonUsefulTokens}, we observe that for Python the dark red curves peak at the initial tokens, indicating a high probability that correct solutions appear at the beginning of the generated sequence and typically end between the 100th and 200th token. However, for models such as \texttt{Qwen2.5-3B-Coder-Instruct}, \texttt{CodeQwen1.5-7B}, and \texttt{Qwen3-4B}, the longer tail indicates that correct solutions can appear later, typically ending around token 400. In Figure~\ref{fig:javaUsefulTokens}, we observe a slightly different spiked shape, but it remains skewed toward the left side of the plot, indicating that correct solutions still tend to appear early in the generated sequence. A slightly lower peak can be explained by the presence of javadoc text at the beginning of the generated functions.
\par Ideally, the histogram bars should align with the peaks of the dark red curve, indicating that the generated solutions consist primarily of tokens that belong to the correct solution. A good example is \texttt{Qwen2.5-7B-Coder} for Python. On average, the correct solution is completed by around the 100th token, and the histogram bars are concentrated within the 0–100 token range. This suggests that the model generates the solution with minimal or no extraneous tokens. Other successful examples for Python include \texttt{CodeGemma-7B}, \texttt{CodeQwen1.5-7B}, \texttt{NextCoder-7B}, \texttt{Phi3.5-4B}, and \texttt{Phi4-4B}. The rest of the models show babbling behaviour when generating Python code. For example, \texttt{CodeLlama-7B} frequently produces outputs around 1000 tokens, even though the correct solution typically appears within the first 0–200 tokens. Tokens generated beyond that point not only contribute no useful output but also waste time and energy, since they are ultimately discarded during post-processing. For Java, even more models exhibit noticeable babbling behavior, such as \texttt{Deepseek-6.7B}, \texttt{Phi4-4B}, and \texttt{Qwen3-4B}. For these models, the red curve and the histograms show little overlap, indicating that the models generate beyond the point of producing the correct solution. Models such as \texttt{Qwen2.5-3B-Coder-Instruct}, \texttt{Qwen2.5-7B-Coder}, and \texttt{CodeQwen1.5-7B},  also display babbling behavior, though it is less pronounced than in the models mentioned earlier. While there is some overlap with the correct solution, these models still generate a substantial number of outputs, approaching the maximum token limit. 
\par The main takeaway from this preliminary experiment is that the correct solution typically appears at the beginning of the generated sequence, after which models tend to start babbling. For models such as \texttt{Qwen2.5-3B-Coder-Instruct}, \texttt{Deepseek-6.7B}, and \texttt{Qwen3-4B}, this tendency occurs in both languages. In contrast, \texttt{CodeLlama-7B} exhibits babbling only for Python, not for Java, while \texttt{Phi4-4B} shows the opposite pattern. So, it shows that any model can potentially exhibit a babbling behavior and that it may be language-specific.

\section{Babbling Suppression}

The method of \textit{babbling suppression} requires two inputs: (i) a prompt that asks an LLM to generate a function and (ii) a set of automated tests that can be used to check the behavior of the generated function. Optionally, a set of libraries on which the generated program or the corresponding tests depend can also be provided. The general idea can be described in three main steps, which aim to answer three questions regarding the generated code: 
\begin{enumerate}
    \item \textbf{Form checking units}: \textit{Is it plausible?}
    \item \textbf{Check well-formedness}: \textit{Is it feasible?}
    \item \textbf{Check functionality}: \textit{Is it acceptable?}
\end{enumerate}
A sequence of generated tokens is \textit{plausible} if it could be correct according to language rules, e.g., a function starts with \texttt{def} in Python and the curly braces match in Java. We say that it is also \textit{feasible} if it is syntactically correct and, for statically-typed languages, it typechecks. Finally, the generated program is \textit{acceptable} if it passes all the provided tests. If an acceptable program has been generated, then the generation process can stop early. However, if the answer to any of the questions above is \textbf{No}, generation continues until all criteria are satisfied or an end-of-generation token is produced. Algorithm~\ref{alg:bs} presents a more detailed description of the method. Section~\ref{sec:overview} describes the three steps of the method in more depth and  Section~\ref{sec:implementation} shows how the proposed approach can be instantiated for the generation of Python and Java programs.

\subsection{Method Overview}\label{sec:overview}
\par \textbf{Form checking units.} In this work, a checking unit is a sequence of tokens that can potentially be syntactically and semantically well-formed, i.e., it can plausibly correspond to a function in a target programming language. This means that it is correctly delimited as a function but its internals are not checked at this step. We check for plausibility first, instead of directly verifying syntax, to reduce the overhead of performing more expensive checks for correct syntax and typing. Ideally, a checking unit  corresponds to a function whose behavior can be verified against the provided test cases. 
\par The identification of checking units depends on the programming language of the generated code and the syntactic delimiters used to define these units (Section~\ref{sec:implementation}). In Algorithm~\ref{alg:bs}, it happens in line 5, which is responsible for breaking the generated sequence of tokens in subsequences bookended by function delimiters. 

The proposed algorithm checks whether a checking unit has been encountered for every generated token. However, it can also be adapted for a specific language. For example, our instantiation for Python does this whenever an end-of-line character is encountered, to reduce the number of checks. 
The decision of when identify checking units must balance two objectives: it should be frequent enough to prevent the model from generating code far beyond a single checking unit, thereby wasting computational resources, yet infrequent enough to avoid excessive checks and unnecessary computational overhead. If a generated token sequence can be successfully transformed into a checking unit, we can proceed by verifying whether it is well-formed.
\par \textbf{Check well-formedness.} At this stage, we evaluate whether the checking unit is well-formed according to the rules of the target language. This comprises verifying the syntax of the generated code and, depending on the language, also performing type checking and name resolution, which can be observed in line 9 of Algorithm~\ref{alg:bs}. 
\par If the checking unit satisfies these well-formedness criteria, it is possible to proceed to the next step. From this point onward, such a checking unit is referred to as a function. However, if the verification fails, we either continue generation or discard the corresponding checking unit. As shown in lines 1,8 , and 16, a dedicated list is maintained for discarded checking units. Discarding a checking unit means that if exactly the same checking unit is encountered again during generation, it will not be forwarded to the subsequent evaluation step. This check is performed in line 8. A checking unit is discarded only when the detected error cannot be resolved through further token generation. For instance, if the error arises from type checking or other structural inconsistencies, additional tokens cannot correct the malformed function. In such cases, the checking unit is permanently discarded. However, if a function is missing a variable or a method, it is not discarded, as the required elements may be introduced in subsequently generated tokens. The call to function \textit{should\_discard()} in line 16 highlights this distinction. It is also important to note that a discarded checking unit may still form a prefix of another checking unit, which is treated independently of the previously discarded one.
\par \textbf{Check functionality.} At this stage, the generated program is executed against the provided test cases as shown in line 11. If all test cases pass, the generation process terminates, as the function satisfies the expected behavior. If the test cases fail, generation continues until either (i) the LLM produces a function that passes the tests, or (ii) the maximum output length or end-of-sequence token is reached.
\bsalgorithm
\begin{figure*}[t]
    \centering
    \includegraphics[width=\linewidth]{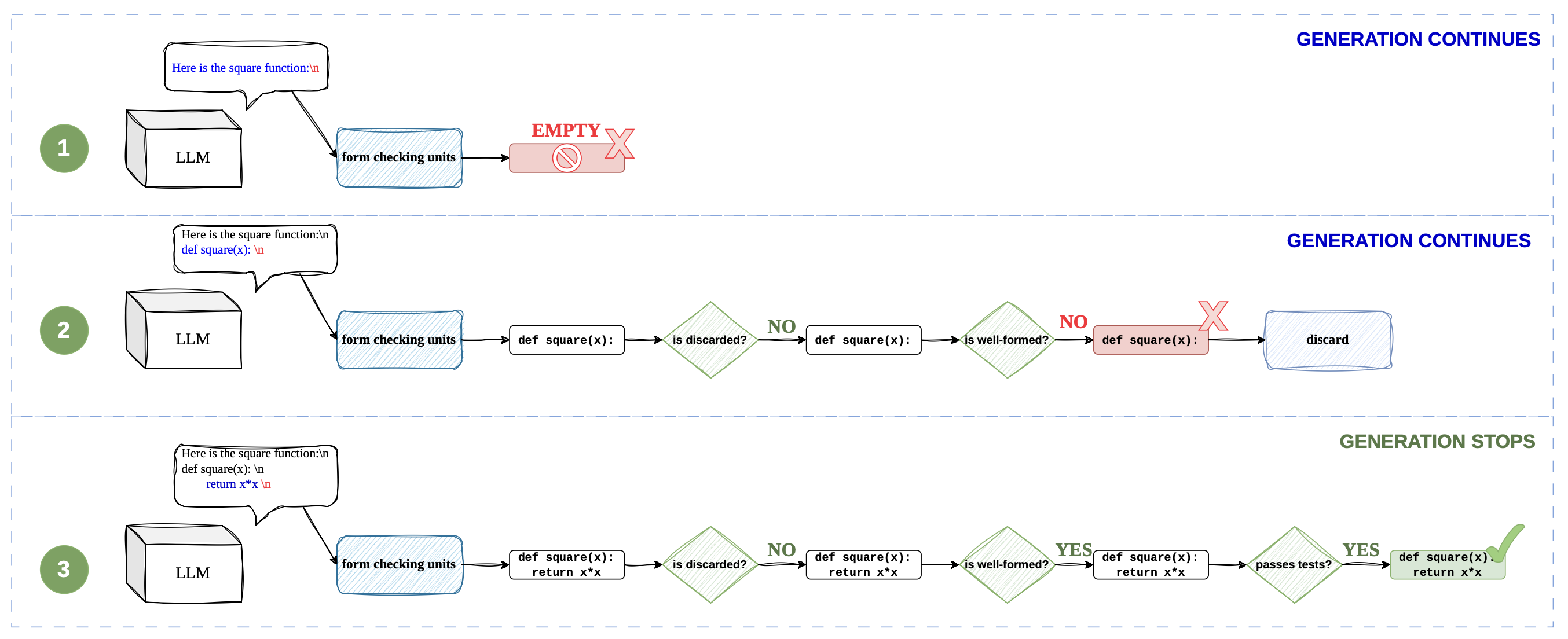}
    \caption{\small
    Step-by-step example of applying \textit{babbling suppression} to an LLM-generated Python function that computes the square of a number. For the generated output: blue indicates a newly generated line, and the line delimiter is highlighted in red. The generated output is shown with a white background, which turns red (with a cross) if a check fails and green (with a check mark) if all tests pass. Blue boxes represent actions, while diamonds denote checks and the corresponding outcomes from the decision.}
    \label{fig:bsexample}
\end{figure*}
\subsection{Instantiating for Python \& Java}\label{sec:implementation}

In this section, we highlight the aspects of the proposed approach that need to be instantiated according to the target programming language, using Python and Java as examples. 
\par \textbf{Form checking units.} Forming checking units (line 5 of Algorithm~\ref{alg:bs}) requires determining the start and end of a function. In Python, a function signature typically has the following form \texttt{def identifier():}, optionally including type annotations and parameters, but it must include the \texttt{def} keyword followed by the function name. In Java, by contrast, function (method) signatures consist of a (possibly empty) list of modifiers, e.g., \texttt{public}, \texttt{private}, \texttt{static}, etc., an explicit return type, which can be a single identifier, a fully-qualified name, or the \texttt{void} keyword, a method name, and parentheses, possibly enclosing a list of parameters. Additional information regarding throwing of exceptions and usage of generics may also be present. 

After identifying the start of the checking unit, it's necessary to find its end. In Python, considering only top-level functions, the body of a function is indicated by indentation: every line following the function signature is expected to be indented, either at the same or a deeper level, and any line that is not indented is considered outside the function body. When such a non-indented line is encountered this marks the end of the checking unit. Java uses curly braces to delimit blocks. This means that the first opening curly brace after the function signature marks the start of the function body and the first closing curly brace at the same level, i.e., when an equal number of opening and closing curly braces is found, marks its end. 

\par \textbf{Check well-formedness.} The two languages also differ in how they verify well-formedness. For Java, we rely on the compiler to detect syntactic and static semantics errors during compilation. For Python, we use the built-in \texttt{compile} function, which attempts to compile a string into an executable code and raises an exception if syntax errors are detected. 

\par During the syntax-checking stage, a checking unit may be discarded depending on the type of error detected. For Python the \texttt{compile} function detects errors such as invalid syntax, missing parentheses, incorrect indentation, and malformed statements. However, it does not detect issues such as references to undefined variables, which are only reported during execution. Thus, if a Python function fails the syntax check, it is discarded, as such errors generally indicate malformed code that cannot be corrected by continuing generation and requires regeneration from scratch. In contrast, the Java compiler can provide more detailed information, including errors related to missing identifiers or unresolved method calls. Syntactically incorrect checking units are discarded just like with Python, as are checking units exhibiting typing errors such as a string being assigned to an integer variable. However, errors stemming from unknown identifiers do not result in the checking unit being discarded, since missing elements may be introduced through subsequent token generation, e.g., an auxiliary function may be generated later. Function \texttt{should\_discard()} in line 15 of Algorithm~\ref{alg:bs} is responsible for verifying these cases. 

\par \textbf{Check functionality.} The execution of test cases is straightforward for both languages. In Python, test cases are appended to the generated file as \texttt{assert} statements placed after the function definition. In Java, we generate a class \texttt{Problem} containing the candidate method, and include \texttt{assert} statements within the \texttt{main} method to invoke the function and check its correctness against the expected outputs. 
\par \textbf{Dependencies.} If \texttt{import} statements are generated, they are included in the final executable file. However, only libraries from the standard Java or Python distributions, or those explicitly provided beforehand, can be used during execution. Consequently, any external dependencies must be made available in advance. For Python, this requires that the necessary libraries be pre-installed in the execution environment. For Java, external dependencies must be supplied as JAR files beforehand so they can be included in the classpath during compilation and execution.
\subsection{Step-by-step Example}
Figure~\ref{fig:bsexample} illustrates a step-by-step example of how babbling suppression is applied during LLM generation. For this example, the LLM is tasked with generating a Python function that computes the square of a number. To reduce the complexity of the example, instead of applying the algorithm every time a token is generated, we apply it each time a new line is encountered, reducing the number of steps needed to achieve a successful completion. Since each step corresponds to the generation of a new line, the most recently generated line is highlighted in blue, while previously generated lines are shown in black. The end-of-line token is shown in red to indicate that it triggers the checks in the algorithm. The LLM output is displayed in a bubble above the labeled box.
\par In the first step, we observe that encountering \texttt{\textbackslash n} triggers the formation of a checking unit, matching line 5 of Algorithm~\ref{alg:bs}. However, since the LLM generates a natural language sentence that does not resemble a function, the formation of the checking unit results in an empty list. Hence, the remaining checks cannot be applied, and generation continues.
\par In the second step, a new line is generated and highlighted in blue. After forming a checking unit, we obtain \texttt{def square(x):}, which corresponds to a function signature. Next, we check whether this checking unit has been previously discarded, matching line 8 of the algorithm, due to failing well-formedness checks. Since it has not, we proceed with the well-formedness check on line 9, which fails because a syntactically correct function requires an indented block after the function signature. As a result, the checking unit is added to the discarded list by the \texttt{should\_discard()} function on line 15 of the algorithm. This checking unit is added because a syntactically incorrect checking unit in Python cannot be corrected by subsequent token generation.
\par In the third step, a new line is generated, and we attempt to form a checking unit, which results in a complete \texttt{square()} function. Although this checking unit shares a prefix with a previously discarded unit, it is treated independently. Hence, when we check the discarded list, the result is negative, as this exact checking unit has not been encountered before. After passing the well-formedness check, we proceed to run tests on this function, corresponding to line 11 of the algorithm. Since all tests pass successfully, generation can be stopped. If the tests were to fail, generation would continue until the model produces the end-of-generation token or reaches the maximum output length.

\section{Experimental design}
To evaluate performance of BS, we focus on the following research questions:
\begin{itemize}
    \item \textbf{\textit{RQ1:}} How effective is BS in truncating LLM outputs while preserving the accuracy of generated code?
    \item \textbf{\textit{RQ2:}} To what extent can BS reduce the energy consumption of LLM inference in code generation tasks?
    \item \textbf{\textit{RQ3:}} Does BS introduce computational overhead, and if so, how significant is it relative to the achieved savings?
\end{itemize}
\par Since the goal of BS is to eliminate excessive tokens to reduce both developer post-processing effort and energy consumption, it is not intended to negatively impact model accuracy. Therefore, \textit{RQ1} and \textit{RQ2} are designed to evaluate these aspects directly. \textit{RQ3} focuses on evaluating the potential overhead introduced by integrating the algorithm along the generation process, as BS performs checks on intermediate outputs, which may results in additional computational cost.
\subsection{Benchmarks \& LLMs}
\par \textbf{Benchmarks.} For benchmark selection, we defined two key requirements: support for function-level code generation and the availability of test cases or input–output pairs. Based on these criteria, we selected HumanEval~\cite{humanEval} and MBPP~\cite{austin2021programsynthesislargelanguage} for Python, and HumanEval Java~\cite{10.1109/TSE.2023.3267446} and APPS~\cite{hendrycks2021measuringcodingchallengecompetence} for Java. We focus on these two languages because they demonstrate notable differences: Python is interpreted and dynamically typed with a concise syntax, while Java is compiled and statically typed. It also enforces more structured code with syntactically-delimited scopes. These differences allow us to demonstrate that our approach generalizes across diverse programming paradigms.
\par For HumanEval, MBPP, and HumanEval Java, we include the problem description and function signature in the prompt. All three benchmarks provide pre-written tests: HumanEval and HumanEval Java contain on average 7–8 unit tests per problem, while MBPP includes around 3. In terms of size, HumanEval contains 164 problems, MBPP - 256, and HumanEval Java - 157. APPS is originally a Python benchmark and provides only problem descriptions and input–output pairs. To incorporate it into our evaluation, we adapt it to Java by constructing function signatures and deriving 5–7 unit tests per problem from the provided input–output pairs. Our adapted version of APPS contains 100 problems. 
\par \textbf{LLMs.} Since our study focuses on code generation from natural language, we selected instruction-based coding LLMs. Furthermore, the models were also chosen based on their HumanEval accuracy, as reported in the Big Code Models Leaderboard~\cite{BigCodeLeaderboard}, while also taking into account our hardware constraints, as all models needed to run locally for energy measurements. The LLMs were loaded via HuggingFace \texttt{transformers 4.56.1} and run using PyTorch 2.8.0. For all experiments, we used a top-p value of 0.95 and a temperature of 0.1 to ensure consistent and comparable conditions across models. Given that our evaluation uses the pass@1 metric, a low temperature was chosen to produce more deterministic outputs. Lastly, all prompts were standardized to ensure that each model received identical input.
\tableResultsEnergyAccuracy
\subsection{Baseline \& Evaluation}
\par \textbf{Baseline.} For the baseline, we use a standard code generation setup in which the instruction and function signature are provided in the prompt. The maximum output length is set to 1000 tokens, giving enough space to observe whether models attempt to reach this limit. Outputs are post-processed for pass@1 evaluation, ensuring that only the generated function is assessed while ignoring any excessive or irrelevant text. 
\par \textbf{Evaluation.} For evaluation, we compare the results of the baseline with the results obtained using code generation with BS. The algorithm is implemented as a \texttt{StoppingCriteria} that is passed to the \texttt{generate()} function in \texttt{transformers}. This criterion is evaluated at each token generation step, and generation is terminated once the condition is satisfied. To compare the baseline and BS, we measure output length, total GPU energy consumption, GPU energy per token, and accuracy. Output length is measured in the number of output tokens. Total GPU energy is measured in Joules, along with GPU energy per token, which is computed as total energy divided by the number of generated tokens. Accuracy is evaluated using pass@1, defined as the proportion of problems for which the generated solution passes all tests
\subsection{Experimental setup \& Energy measurements}
\par \textbf{Experimental setup.} The experiments were carried out using JupyterLab
 with IPython 8.21.0 and Python 3.10.12. Java programs were compiled using javac 11.0.29 and executed with OpenJDK 11.0.29 on Ubuntu 22.04. The tests were supplied as \texttt{assert} statements, which were enabled at runtime using the \texttt{-ea} flag. We utilized a GPU cluster with NVIDIA A10 GPU with 24GB of memory. 
\par \textbf{Energy measurements.} All runs were performed on a single machine, with no other user jobs running at the same time, ensuring no interference from other processes. Prior to each inference run, the system was checked to confirm it was idle. GPU power consumption was monitored using pyNVML~\cite{pynvml}, a Python interface to the NVIDIA Management Library~\cite{nvidia:2025}. Measurements were sampled at 10 Hz to capture fine-grained energy usage during inference. The energy consumption is computed as the product of average power and execution time.


\section{Results \& Discussion}
In this section, we present our main findings. Section~\ref{sec:acc_energy} analyzes the impact of BS on generation accuracy and total GPU energy consumption, addressing \textit{RQ1} and \textit{RQ2}, respectively. Section~\ref{sec:overhead} focuses on \textit{RQ3} by examining the overhead introduced by the algorithm.
\subsection{Impact on accuracy \& energy}
\label{sec:acc_energy}
\tableEpT
Table~\ref{tab:EnergyAccuracy} shows that, in most cases, introducing BS does not affect accuracy. Only two models deviate from the baseline: \texttt{Qwen2.5-Coder-3B} and \texttt{Qwen3-4B}. Observed improvements and decreases in accuracy after applying BS can be attributed to the inherent non-determinism of the models, as BS operates externally to the model and does not interfere with decoding.
For example, for the Python benchmarks,\texttt{Qwen3-4B} with BS exhibits 6\% higher accuracy than the baseline for HumanEval and 3\% lower accuracy than the baseline for MBPP. We manually analyzed the solutions generated by \texttt{Qwen3-4B} to investigate differences in test pass rates between the baseline and BS. Even with the same low temperature and high top-p settings, the model produced different outputs for the same prompt across runs. In some cases, the baseline produced a correct solution, while in others, the run with BS did. Accuracies were not statistically significantly different for any of the models or benchmarks, based on Two-proportion Z-test, even setting the alpha to a relatively high value, e.g., 0.1. 

\par Table~\ref{tab:EnergyAccuracy} presents the mean energy consumption of model inference, together with the average output length, for code generation with BS compared to the baseline. The table also includes a $\Delta$ column, showing the difference between the baseline and BS, with green arrows indicating savings and red arrows indicating increased cost. We highlight in boldface the cases where the difference was statistically significantly different, based on the Mann-Whitney U test ($\alpha = 0.005$). We observe that BS reduces energy consumption for both Python and Java, with savings of up to 65\% for Python and up to 62\% for Java.

\par From Figure~\ref{fig:pythonUsefulTokens}, we identify four models that show babbling behavior in Python: \texttt{Qwen2.5-3B-Coder-Instruct}, \texttt{CodeLlama-7B}, \texttt{Qwen3-4B}, and \texttt{Deepseek-6.7B}. Table~\ref{tab:python} shows that BS achieves mean energy savings between 21\% and 58\% on the HumanEval benchmark and between 19\% and 65\% on MBPP across all these models. The magnitude of the savings can vary, depending on the model’s accuracy. This is because the algorithm only terminates generation once the test cases pass, which requires the LLM to be capable of producing a correct solution. Since \texttt{CodeLlama-7B} achieves only 33\% accuracy on HumanEval, the overall energy savings for the full benchmark are limited to 21\%. This is because the mean is calculated over all data points in the benchmark, not just those that pass the tests. For MBPP, \texttt{CodeLlama-7B} achieves a higher accuracy of 41\%, resulting in greater energy savings of 39\%. In the case of \texttt{Qwen3-4B}, the accuracy on HumanEval is 76\%, yet the energy savings are only 30\%. This is explained by the correct solution often appearing later in the generated sequence. As shown in Figure~\ref{fig:pythonUsefulTokens}, \texttt{Qwen3-4B} exhibits a long and dense tail, indicating that correct solutions can occur toward the end of the sequence.
\par With BS, we also achieved energy savings for the other models, including those that did not exhibit obvious babbling behavior. For example, baseline \texttt{Phi4-4B} produced only 182 tokens on average per solution for HumanEval, but the use of BS reduced that number to 121, i.e., 34\% fewer tokens.  Notwithstanding, for two models, \texttt{Qwen2.5-Coder-7B} and \texttt{CodeGemma-7B}, the algorithm resulted in additional energy costs. Although the number of output tokens was still reduced by 2\% and 6\%, respectively, the overhead introduced by BS caused the mean energy consumption to increase by 8\% and 9\%. Figure~\ref{fig:pythonUsefulTokens} shows that these two models produced outputs that were already short, containing just the tokens needed for a correct solution. Hence, there was little room for further reduction.
\par For Java, Figure~\ref{fig:javaUsefulTokens} shows that a larger number of models babble, with histograms skewed toward the right side of the plots, indicating a substantial number of solutions that are one thousand-tokens long. This pattern is reflected in Table~\ref{tab:java} for HumanEvalJava, where BS achieves mean energy savings ranging from 20\% to 62\% across all models. The only model showing less babbling is \texttt{NextCoder-7B}, which had the lowest energy savings of 20\%.
\par The true limitations of BS become apparent on the APPS benchmark. Model accuracy on this benchmark ranges from 1\% to 20\%, leaving little room for truncation. In the original work by Hendrycks et al.~\cite{hendrycks2021measuringcodingchallengecompetence}, the APPS benchmark was shown to be challenging even for large language models. For example, GPT-Neo solves only about 20\% of the introductory-level problems. So due to low accuracies, BS shows limited potential for energy savings. Only five out of ten models had their outputs truncated, and only two models achieved energy savings: 1\% for \texttt{Qwen3-4B} and 7\% for \texttt{Qwen2.5-Coder-3B}. The remaining models incurred additional energy costs of 1\% to 10\% due to the overhead introduced by BS.

\par The use of BS achieved reduction in the number of tokens consistently across the benchmarks, considering the two languages. Even for APPS, where the very low accuracy imposes an intrinsic limitation to the achievable reduction, BS resulted in five models producing fewer tokens than the corresponding baselines.
\begin{summary}
{\footnotesize 
\textbf{Summary.} 
\hl{The results indicate that BS does not negatively affect model accuracy. Furthermore, it can achieve energy savings for all models in both Python and Java, reaching up to 65\% and 62\%, respectively. The largest savings occur for models that tend to babble after generating a correct solution early in the sequence, such as \texttt{Deepseek-6.7B} and \texttt{Qwen2.5-Coder-3B}. Energy savings are also directly tied to model accuracy. If a model cannot produce a correct solution, then BS is unable to truncate the output. This can lead to additional energy consumption due to the overhead introduced by BS. Finally, the use of BS resulted in fewer output tokens across all the models and benchmarks, with few exceptions.}
}
\end{summary}
\subsection{Overhead}\label{sec:overhead}
\label{sec:overhead}
\par Table~\ref{tab:eptAll} shows GPU energy per token across models, comparing BS with the baseline. BS introduces overhead in most cases, ranging from 2\% to 24\%, except for \texttt{DeepSeek-6.7B} and \texttt{Qwen2.5-Coder-3B} on HumanEval Java, which can be attributed to substantial output truncation. For the former the average output decreases from 988 to 489 tokens and for the latter from 608 to 242 tokens. As shown by Solovyeva et al.~\cite{solovyeva2026greenaidecodingenergy}, the energy per token increases with each newly generated token. Thus, earlier tokens are less energy-intensive, so truncation lowers the average energy per token. This does not eliminate overhead, but shows that the reduction in output length outweighs it.
\tableOverhead
\par For a more detailed analysis of the overhead, we refer to Table~\ref{tab:overhead}, which presents data for three models on HumanEval: \texttt{CodeLlama-7B}, \texttt{Qwen2.5-Coder-7B}, and \texttt{Phi4-4B}. \texttt{CodeLlama-7B} was selected to illustrate overhead in a low-accuracy scenario, \texttt{Qwen2.5-Coder-7B} to show the overhead when there is little or no potential for output truncation, and \texttt{Phi4-4B} represents an average case with acceptable accuracy and a possibility for output truncation. The table presents several previously discussed metrics, including output length, energy per token, and GPU energy. It also provides: (i) average time per token, representing the time to generate a single token; (ii) BS time, indicating the time spent on BS-related tasks, such as checking well-formedness and running tests; (iii) total time for a single inference; (iv) average GPU power; and (v) the average number of test suite executions by BS per inference.
\par Table~\ref{tab:overhead} shows that the increase in energy per token corresponds to an increase in time to token, remaining around 5–6ms across all three models. Additional time is attributed to running BS, which performs checks on the intermediate output. A single iteration of BS can take between 127ms and 196ms, with its duration depending on the complexity of the function and the number of tests that must be executed. The duration of a single test suite execution directly impacts the overall runtime of BS. If a model generates a solution with a non-terminating infinite loop, the test would run indefinitely. So, BS enforces a 10-second timeout per test. The table also reports the average number of times BS executes tests, which averages around 3 executions per inference run.
\par To offset the overhead, the overall savings must exceed the additional costs. So, despite the overhead, code generation with BS can reduce the total inference time. For example, \texttt{CodeLlama-7B} and \texttt{Phi4-4B} achieve average reductions of 18\% and 23\%, respectively, even in the presence of this overhead. For \texttt{Qwen2.5-Coder-7B}, where the output is truncated by only 2\%, the overhead leads to a 21\% increase in total runtime. This suggests that the model already performs well on HumanEval, producing concise and correct solutions. However, the benefits of BS become evident for models that tend to overproduce. The same model exhibits substantial babbling on HumanEval Java, where BS reduces output length by 52\%, resulting in 45\% energy savings. This indicates that a model’s behavior is not consistent across languages: even if it does not babble in one language, it may still do so in another.
\par Lastly, the Table~\ref{tab:overhead} also shows that the average GPU power is consistently higher for the baseline than for BS. This is explained by BS operating on the CPU due to token decoding with the tokenizer and test executions. As a result, parts of the inference process shift from GPU to CPU, leaving the GPU intermittently idle and lowering its average power consumption. We can treat BS time as CPU time, since it runs on CPU, and estimate its contribution to the total runtime of code generation with BS. For a single test execution, BS accounts for approximately 1\% to 4\% of the total duration. Considering the average of three test executions per inference (as shown in Table~\ref{tab:overhead}), this contribution increases to roughly 2\%–10\%, indicating that the majority of the inference workload remains on the GPU.

\begin{summary}
{\footnotesize 
\textbf{Summary.} 
\hl{The results show that BS introduces overhead, with energy per token increasing by 2\%–24\% compared to the baseline. This overhead stems from the additional checks and test executions performed to verify correctness before interrupting generation, and is therefore influenced by both the duration and number of test cases. Despite this, a single BS iteration adds only 127–196 ms to an inference run, contributing just 1\%–4\% to the total runtime. On average, BS runs about three iterations per inference. Lastly, it also shifts part of the workload to the CPU, slightly reducing average GPU power. However, at least 90\% of the workload still remains on the GPU.}
}
\end{summary}

\section{Threats to validity}\label{sec:threats}

\par \textbf{Construct validity.} A potential threat to construct validity is that CPU energy usage during BS is not measured directly. 
First, we do not have direct access to CPU energy measurements as we do not have physical access to the server nor root privileges. Second, as shown in Section~\ref{sec:overhead}, these results would likely not be meaningful, since BS results in multiple quick transitions between GPU and CPU. Each one of them lasts often less than one millisecond and interfaces such as Intel RAPL do not provide a high enough sampling rate to reliably measure such short executions~\cite{Khan:2018:RAE}. So, using time as a proxy is potentially more reliable, as contemporary computers can measure time at a granularity of nanoseconds. Previous work~\cite{alizadeh2025languagemodelssoftwaredevelopment} has shown that the correlation between inference time and energy for LLM-based code generation is high.
\par \textbf{Internal validity.} A threat to internal validity arises from hardware constraints, as the experiments were conducted in a shared environment where exclusive GPU access could not be granted. This introduces the possibility that activity from other users on the same GPU had a chance to influence energy measurements. However, to mitigate this risk, GPU utilization was monitored prior to each inference run, and execution was only started when a single active process was detected, thereby ensuring exclusive access.
\par \textbf{External validity.} A threat to external validity is that our findings are limited to two programming languages, Python and Java. Therefore, we cannot guarantee that babbling occurs in other languages or that BS generalizes to different contexts. Nevertheless, these languages were selected due to their substantial differences: Python is dynamically typed with indentation-based blocks, whereas Java is statically typed and relies on explicit structure. These differences would help broaden coverage, although generalizability beyond these languages would need more analysis.

\section{Conclusion}
In this work, we examined the problem of \textbf{“babbling”} in LLM-based code generation, defined as the generation of excessive tokens that are typically removed during post-processing or ignored by developers. We proposed Babbling Suppression (BS), a simple yet effective approach that integrates test execution into the generation process to terminate generation once a correct solution is reached. Our empirical evaluation with ten models and four benchmarks across two languages demonstrates that babbling is a common phenomenon, particularly in Java. BS consistently reduces unnecessary token generation without compromising correctness. The results show energy savings of up to 65\% for Python and 62\% for Java, while maintaining low overhead, leading to net efficiency gains in 73\% of the cases. Overall, BS provides a practical and model-agnostic solution for improving the efficiency of AI-assisted code generation.

\section{Data Availability Statement}
The replication package is available on Zenodo~\cite{replicationPackage}. It includes Python notebooks for running all models across the benchmarks, both for the baseline and with BS. The package also contains the raw energy measurements used in our analysis, as well as scripts for generating all graphs and figures. The package is provided as a compressed archive and should be unzipped after downloading. It includes a README file with more detailed information.  
\bibliographystyle{ACM-Reference-Format}
\bibliography{sample-base}

@article{Khan:2018:RAE,
author = {Khan, Kashif Nizam and Hirki, Mikael and Niemi, Tapio and Nurminen, Jukka K. and Ou, Zhonghong},
title = {RAPL in Action: Experiences in Using RAPL for Power Measurements},
year = {2018},
issue_date = {June 2018},
publisher = {Association for Computing Machinery},
volume = {3},
number = {2},
issn = {2376-3639},
journal = {ACM Trans. Model. Perform. Eval. Comput. Syst.},
month = mar,
articleno = {9},
numpages = {26},
}

@INPROCEEDINGS{Tang:2024:DBV,
  author={Tang, Ningzhi and Chen, Meng and Ning, Zheng and Bansal, Aakash and Huang, Yu and McMillan, Collin and Li, Toby Jia-Jun},
  booktitle={2024 IEEE Symposium on Visual Languages and Human-Centric Computing (VL/HCC)}, 
  title={Developer Behaviors in Validating and Repairing LLM-Generated Code Using IDE and Eye Tracking}, 
  year={2024},
  volume={},
  number={},
  pages={40-46},
}

@misc{Economist:2025:OLM,
      title={{OpenAI}'s latest model will change the economics of software}, 
      author={{The Economist}},
      year={2025},
      month = {January}, 
      note={\url{https://www.economist.com/business/2025/01/20/openais-latest-model-will-change-the-economics-of-software}}, 
}

@misc{Economist:2025:WOE,
      title={Will {OpenAI} ever make real money?}, 
      author={{The Economist}},
      year={2025},
      month = {May}, 
      note={\url{https://www.economist.com/business/2025/05/15/will-openai-ever-make-real-money}}, 
}

@misc{Zhang:2024:VVD,
      title={Verbosity $\neq$ Veracity: Demystify Verbosity Compensation Behavior of Large Language Models}, 
      author={Yusen Zhang and Sarkar Snigdha Sarathi Das and Rui Zhang},
      year={2024},
      eprint={2411.07858},
      archivePrefix={arXiv},
      primaryClass={cs.CL},
      url={https://arxiv.org/abs/2411.07858}, 
}

@ARTICLE{GenAISoftware,
  author={Ebert, Christof and Louridas, Panos},
  journal={IEEE Software}, 
  title={Generative AI for Software Practitioners}, 
  year={2023},
  volume={40},
  number={4},
  pages={30-38},
  doi={10.1109/MS.2023.3265877}}

@misc{pynvml,
  author       = "{PyPI Contributors}",
  title        = "{pynvml: Python bindings for NVML}",
  year         = "2024",
  url          = "https://pypi.org/project/pynvml/",
  note         = "Accessed: 2025-10-23"
}

@misc{nvidia:2025,
      title={{NVIDIA Management Library (NVML)}}, 
      author={{NVIDIA Corporation}},
      year={2025},
      note={https://developer.nvidia.com/management-library-nvml. Last accessed October 22nd, 2025.}
}

@misc{githubCopilot, 
    author     = "GitHub", 
    title      = "GitHub Copilot", 
    year       = "2025", 
    url        = "https://copilot.github.com/",
    note       = "Accessed: 2025-10-23"
}

@misc{claude, 
    author     = "ANTHROPIC PBC", 
    title      = "Claude", 
    year       = "2026", 
    url        = "https://claude.ai/",
    note       = "Accessed: 2026-03-23"
}

@misc{chatGPT, 
    author     = "OpenAI", 
    title      = "ChatGPT", 
    year       = "2025", 
    url        = "https://chat.openai.com/chat",
    note       = "Accessed: 2025-10-23"
}

@misc{babble, 
    author     = "Merriam-Webster", 
    title      = "babbling", 
    year       = "2026", 
    url        = "https://www.merriam-webster.com/dictionary/babbling",
    note       = "Accessed: 2026-01-22"
}

@misc{BigCodeLeaderboard, 
    author     = "BigCode", 
    title      = "", 
    year       = "2025", 
    url        = "https://huggingface.co/spaces/bigcode/bigcode-models-leaderboard",
    note       = "Accessed: 2025-10-23"
}

@article{humanEval,
  author={Mark Chen and Jerry Tworek and Heewoo Jun and Qiming Yuan and Henrique Ponde de Oliveira Pinto and Jared Kaplan and Harri Edwards and others},
  title        = {Evaluating Large Language Models Trained on Code},
  journal      = {CoRR},
  volume       = {abs/2107.03374},
  year         = {2021},
}

@INPROCEEDINGS{alizadeh2025languagemodelssoftwaredevelopment,
  author={Alizadeh, Negar and Belchev, Boris and Saurabh, Nishant and Kelbert, Patricia and Castor, Fernando},
  booktitle={2025 IEEE/ACM 22nd International Conference on Mining Software Repositories (MSR)}, 
  title={Language Models in Software Development Tasks: An Experimental Analysis of Energy and Accuracy}, 
  year={2025},
  volume={},
  number={},
  pages={725-736},
  doi={10.1109/MSR66628.2025.00109}}

@inproceedings{wattsLuccioni,
author = {Luccioni, Sasha and Jernite, Yacine and Strubell, Emma},
title = {Power Hungry Processing: Watts Driving the Cost of AI Deployment?},
year = {2024},
isbn = {9798400704505},
publisher = {Association for Computing Machinery},
address = {New York, NY, USA},
url = {https://doi.org/10.1145/3630106.3658542},
doi = {10.1145/3630106.3658542},
booktitle = {Proceedings of the 2024 ACM Conference on Fairness, Accountability, and Transparency},
pages = {85–99},
numpages = {15},
location = {Rio de Janeiro, Brazil},
series = {FAccT '24}
}

@INPROCEEDINGS{ai-powered-power-hungry,
  author={Solovyeva, Lola and Weidmann, Sophie and Castor, Fernando},
  booktitle={2025 IEEE/ACM Second International Conference on AI Foundation Models and Software Engineering (Forge)}, 
  title={AI-Powered, But Power-Hungry? Energy Efficiency of LLM-Generated Code}, 
  year={2025},
  volume={},
  number={},
  pages={49-60},
  doi={10.1109/Forge66646.2025.00012}}

@article{hallucinationsCodeGen, author = {Zhang, Ziyao and Wang, Chong and Wang, Yanlin and Shi, Ensheng and Ma, Yuchi and Zhong, Wanjun and Chen, Jiachi and Mao, Mingzhi and Zheng, Zibin}, title = {LLM Hallucinations in Practical Code Generation: Phenomena, Mechanism, and Mitigation}, year = {2025}, issue_date = {July 2025}, publisher = {Association for Computing Machinery}, address = {New York, NY, USA}, volume = {2}, number = {ISSTA}, url = {https://doi.org/10.1145/3728894}, doi = {10.1145/3728894}, journal = {Proc. ACM Softw. Eng.}, month = jun, articleno = {ISSTA022}, numpages = {23} }

@INPROCEEDINGS{codeQualityLLM,
  author={Jamil, Mohammad Talal and Abid, Shamsa and Shamail, Shafay},
  booktitle={2025 IEEE/ACM 22nd International Conference on Mining Software Repositories (MSR)}, 
  title={Can LLMs Generate Higher Quality Code Than Humans? An Empirical Study}, 
  year={2025},
  volume={},
  number={},
  pages={478-489},
  doi={10.1109/MSR66628.2025.00081}}

@misc{apsan2025,
      title={Generating Energy-Efficient Code via Large-Language Models -- Where are we now?}, 
      author={Radu Apsan and Vincenzo Stoico and Michel Albonico and Rudra Dhar and Karthik Vaidhyanathan and Ivano Malavolta},
      year={2025},
      eprint={2509.10099},
      archivePrefix={arXiv},
      primaryClass={cs.SE},
      url={https://arxiv.org/abs/2509.10099}, 
}

@inproceedings{10.1145/3643795.3648379,
author = {Rasnayaka, Sanka and Wang, Guanlin and Shariffdeen, Ridwan and Iyer, Ganesh Neelakanta},
title = {An Empirical Study on Usage and Perceptions of LLMs in a Software Engineering Project},
year = {2024},
isbn = {9798400705793},
publisher = {Association for Computing Machinery},
address = {New York, NY, USA},
url = {https://doi.org/10.1145/3643795.3648379},
doi = {10.1145/3643795.3648379},
booktitle = {Proceedings of the 1st International Workshop on Large Language Models for Code},
pages = {111–118},
numpages = {8},
location = {Lisbon, Portugal},
series = {LLM4Code '24}
}

@INPROCEEDINGS{10628428,
  author={Jahić, Jasmin and Sami, Ashkan},
  booktitle={2024 IEEE 21st International Conference on Software Architecture Companion (ICSA-C)}, 
  title={State of Practice: LLMs in Software Engineering and Software Architecture}, 
  year={2024},
  volume={},
  number={},
  pages={311-318},
  doi={10.1109/ICSA-C63560.2024.00059}}

@inproceedings{10.1145/3639477.3643648,
author = {Davila, Nicole and Wiese, Igor and Steinmacher, Igor and Lucio da Silva, Lucas and Kawamoto, Andre and Favaro, Gilson Jose Peres and Nunes, Ingrid},
title = {An Industry Case Study on Adoption of AI-based Programming Assistants},
year = {2024},
isbn = {9798400705014},
publisher = {Association for Computing Machinery},
address = {New York, NY, USA},
url = {https://doi.org/10.1145/3639477.3643648},
doi = {10.1145/3639477.3643648},
booktitle = {Proceedings of the 46th International Conference on Software Engineering: Software Engineering in Practice},
pages = {92–102},
numpages = {11},
location = {Lisbon, Portugal},
series = {ICSE-SEIP '24}
}

@misc{mehditabar2025smartcostlybenchmarkingllms,
      title={Smart but Costly? Benchmarking LLMs on Functional Accuracy and Energy Efficiency}, 
      author={Mohammadjavad Mehditabar and Saurabhsingh Rajput and Antonio Mastropaolo and Tushar Sharma},
      year={2025},
      eprint={2511.07698},
      archivePrefix={arXiv},
      primaryClass={cs.SE},
      url={https://arxiv.org/abs/2511.07698}, 
}

@misc{solovyeva2026greenaidecodingenergy,
      title={Towards Green AI: Decoding the Energy of LLM Inference in Software Development}, 
      author={Lola Solovyeva and Fernando Castor},
      year={2026},
      eprint={2602.05712},
      archivePrefix={arXiv},
      primaryClass={cs.SE},
      url={https://arxiv.org/abs/2602.05712}, 
}

@article{SERGEYUK2025107610,
title = {Using AI-based coding assistants in practice: State of affairs, perceptions, and ways forward},
journal = {Information and Software Technology},
volume = {178},
pages = {107610},
year = {2025},
issn = {0950-5849},
doi = {https://doi.org/10.1016/j.infsof.2024.107610},
url = {https://www.sciencedirect.com/science/article/pii/S0950584924002155},
author = {Agnia Sergeyuk and Yaroslav Golubev and Timofey Bryksin and Iftekhar Ahmed}
}

@article{10.1145/3747588,
author = {Jiang, Juyong and Wang, Fan and Shen, Jiasi and Kim, Sungju and Kim, Sunghun},
title = {A Survey on Large Language Models for Code Generation},
year = {2026},
issue_date = {February 2026},
publisher = {Association for Computing Machinery},
address = {New York, NY, USA},
volume = {35},
number = {2},
issn = {1049-331X},
url = {https://doi.org/10.1145/3747588},
doi = {10.1145/3747588},
journal = {ACM Trans. Softw. Eng. Methodol.},
month = jan,
articleno = {58},
numpages = {72}
}

@ARTICLE{11429547,
author={Kharma, Mohammed F. and Choi, Soohyeon and Alkhanafseh, Mohammad and Mohaisen, David},
journal={ IEEE Transactions on Dependable and Secure Computing },
title={{ Security and Quality in LLM-Generated Code: a Multi-Language, Multi-Model Analysis }},
year={5555},
volume={},
number={01},
ISSN={1941-0018},
pages={1-15},
doi={10.1109/TDSC.2026.3672745},
url = {https://doi.ieeecomputersociety.org/10.1109/TDSC.2026.3672745},
publisher={IEEE Computer Society},
address={Los Alamitos, CA, USA},
month=mar}

@inproceedings{10.1145/3691620.3695470,
author = {Cao, Jialun and Chen, Zhiyong and Wu, Jiarong and Cheung, Shing-Chi and Xu, Chang},
title = {JavaBench: A Benchmark of Object-Oriented Code Generation for Evaluating Large Language Models},
year = {2024},
isbn = {9798400712487},
publisher = {Association for Computing Machinery},
address = {New York, NY, USA},
url = {https://doi.org/10.1145/3691620.3695470},
doi = {10.1145/3691620.3695470},
booktitle = {Proceedings of the 39th IEEE/ACM International Conference on Automated Software Engineering},
pages = {870–882},
numpages = {13},
location = {Sacramento, CA, USA},
series = {ASE '24}
}

@misc{white2025livebenchchallengingcontaminationlimitedllm,
      title={LiveBench: A Challenging, Contamination-Limited LLM Benchmark}, 
      author={Colin White and Samuel Dooley and Manley Roberts and Arka Pal and Ben Feuer and Siddhartha Jain and Ravid Shwartz-Ziv and Neel Jain and Khalid Saifullah and Sreemanti Dey and Shubh-Agrawal and Sandeep Singh Sandha and Siddartha Naidu and Chinmay Hegde and Yann LeCun and Tom Goldstein and Willie Neiswanger and Micah Goldblum},
      year={2025},
      eprint={2406.19314},
      archivePrefix={arXiv},
      primaryClass={cs.CL},
      url={https://arxiv.org/abs/2406.19314}, 
}

@misc{hendrycks2021measuringcodingchallengecompetence,
      title={Measuring Coding Challenge Competence With APPS}, 
      author={Dan Hendrycks and Steven Basart and Saurav Kadavath and Mantas Mazeika and Akul Arora and Ethan Guo and Collin Burns and Samir Puranik and Horace He and Dawn Song and Jacob Steinhardt},
      year={2021},
      eprint={2105.09938},
      archivePrefix={arXiv},
      primaryClass={cs.SE},
      url={https://arxiv.org/abs/2105.09938}, 
}

@misc{austin2021programsynthesislargelanguage,
      title={Program Synthesis with Large Language Models}, 
      author={Jacob Austin and Augustus Odena and Maxwell Nye and Maarten Bosma and Henryk Michalewski and David Dohan and Ellen Jiang and Carrie Cai and Michael Terry and Quoc Le and Charles Sutton},
      year={2021},
      eprint={2108.07732},
      archivePrefix={arXiv},
      primaryClass={cs.PL},
      url={https://arxiv.org/abs/2108.07732}, 
}

@article{10.1109/TSE.2023.3267446,
author = {Cassano, Federico and Gouwar, John and Nguyen, Daniel and Nguyen, Sydney and Phipps-Costin, Luna and Pinckney, Donald and Yee, Ming-Ho and Zi, Yangtian and Anderson, Carolyn Jane and Feldman, Molly Q and Guha, Arjun and Greenberg, Michael and Jangda, Abhinav},
title = {MultiPL-E: A Scalable and Polyglot Approach to Benchmarking Neural Code Generation},
year = {2023},
issue_date = {July 2023},
publisher = {IEEE Press},
volume = {49},
number = {7},
issn = {0098-5589},
url = {https://doi.org/10.1109/TSE.2023.3267446},
doi = {10.1109/TSE.2023.3267446},
journal = {IEEE Trans. Softw. Eng.},
month = jul,
pages = {3675–3691},
numpages = {17}
}

@misc{wang2025codecontestshighqualitytestcase,
      title={CodeContests+: High-Quality Test Case Generation for Competitive Programming}, 
      author={Zihan Wang and Siyao Liu and Yang Sun and Hongyan Li and Kai Shen},
      year={2025},
      eprint={2506.05817},
      archivePrefix={arXiv},
      primaryClass={cs.SE},
      url={https://arxiv.org/abs/2506.05817}, 
}

@article{10.1145/3660810,
author = {Mu, Fangwen and Shi, Lin and Wang, Song and Yu, Zhuohao and Zhang, Binquan and Wang, ChenXue and Liu, Shichao and Wang, Qing},
title = {ClarifyGPT: A Framework for Enhancing LLM-Based Code Generation via Requirements Clarification},
year = {2024},
issue_date = {July 2024},
publisher = {Association for Computing Machinery},
address = {New York, NY, USA},
volume = {1},
number = {FSE},
url = {https://doi.org/10.1145/3660810},
doi = {10.1145/3660810},
journal = {Proc. ACM Softw. Eng.},
month = jul,
articleno = {103},
numpages = {23}
}

@misc{ilager2025greencodelearningoptimizeenergy,
      title={GREEN-CODE: Learning to Optimize Energy Efficiency in LLM-based Code Generation}, 
      author={Shashikant Ilager and Lukas Florian Briem and Ivona Brandic},
      year={2025},
      eprint={2501.11006},
      archivePrefix={arXiv},
      primaryClass={cs.DC},
      url={https://arxiv.org/abs/2501.11006}, 
}

@inproceedings{10.1145/3650212.3680343,
author = {Guo, Lianghong and Wang, Yanlin and Shi, Ensheng and Zhong, Wanjun and Zhang, Hongyu and Chen, Jiachi and Zhang, Ruikai and Ma, Yuchi and Zheng, Zibin},
title = {When to Stop? Towards Efficient Code Generation in LLMs with Excess Token Prevention},
year = {2024},
isbn = {9798400706127},
publisher = {Association for Computing Machinery},
address = {New York, NY, USA},
url = {https://doi.org/10.1145/3650212.3680343},
doi = {10.1145/3650212.3680343},
booktitle = {Proceedings of the 33rd ACM SIGSOFT International Symposium on Software Testing and Analysis},
pages = {1073–1085},
numpages = {13},
location = {Vienna, Austria},
series = {ISSTA 2024}
}

@INPROCEEDINGS{10403378,
  author={Wang, Jianxun and Chen, Yixiang},
  booktitle={2023 IEEE International Conference on Medical Artificial Intelligence (MedAI)}, 
  title={A Review on Code Generation with LLMs: Application and Evaluation}, 
  year={2023},
  volume={},
  number={},
  pages={284-289},
  doi={10.1109/MedAI59581.2023.00044}}

@article{KUHAIL2024103111,
title = {“Will I be replaced?” Assessing ChatGPT's effect on software development and programmer perceptions of AI tools},
journal = {Science of Computer Programming},
volume = {235},
pages = {103111},
year = {2024},
issn = {0167-6423},
doi = {https://doi.org/10.1016/j.scico.2024.103111},
url = {https://www.sciencedirect.com/science/article/pii/S0167642324000340},
author = {Mohammad Amin Kuhail and Sujith Samuel Mathew and Ashraf Khalil and Jose Berengueres and Syed Jawad Hussain Shah}
}

@incollection{REVURI2026115,
title = {Chapter Five - Artificial intelligence (AI) technologies and tools for accelerated software development},
editor = {Pethuru Raj and Mats Agerstam and Pushan Kumar Dutta and B. Sundaravadivazhagan and Gayathri Nagasubramanian},
series = {Advances in Computers},
publisher = {Elsevier},
volume = {141},
pages = {115-159},
year = {2026},
booktitle = {Cloud-native Architecture (CNA) and Artificial Intelligence (AI) for the Future of Software Engineering: The Principles, Patterns, Platforms and Practices},
issn = {0065-2458},
doi = {https://doi.org/10.1016/bs.adcom.2025.07.001},
url = {https://www.sciencedirect.com/science/article/pii/S0065245825001111},
author = {Jaswanth Revuri and Rakesh Kumar Sakthivel and Gayathri Nagasubramanian}
}

@article{HUSEIN2025103917,
title = {Large language models for code completion: A systematic literature review},
journal = {Computer Standards \& Interfaces},
volume = {92},
pages = {103917},
year = {2025},
issn = {0920-5489},
doi = {https://doi.org/10.1016/j.csi.2024.103917},
url = {https://www.sciencedirect.com/science/article/pii/S0920548924000862},
author = {Rasha Ahmad Husein and Hala Aburajouh and Cagatay Catal}
}

@article{10.1145/3801158,
author = {Cordeiro, Jonathan and Noei, Shayan and Zou, Ying},
title = {An Empirical Study on the Code Refactoring Capability of Large Language Models},
year = {2026},
publisher = {Association for Computing Machinery},
address = {New York, NY, USA},
issn = {1049-331X},
url = {https://doi.org/10.1145/3801158},
doi = {10.1145/3801158},
note = {Just Accepted},
journal = {ACM Trans. Softw. Eng. Methodol.},
month = mar
}

@inproceedings{10.1145/3613904.3641936,
author = {Mozannar, Hussein and Bansal, Gagan and Fourney, Adam and Horvitz, Eric},
title = {Reading Between the Lines: Modeling User Behavior and Costs in AI-Assisted Programming},
year = {2024},
isbn = {9798400703300},
publisher = {Association for Computing Machinery},
address = {New York, NY, USA},
url = {https://doi.org/10.1145/3613904.3641936},
doi = {10.1145/3613904.3641936},
booktitle = {Proceedings of the 2024 CHI Conference on Human Factors in Computing Systems},
articleno = {142},
numpages = {16},
location = {Honolulu, HI, USA},
series = {CHI '24}
}

@inproceedings{10.1145/3691620.3695527,
author = {Mathews, Noble Saji and Nagappan, Meiyappan},
title = {Test-Driven Development and LLM-based Code Generation},
year = {2024},
isbn = {9798400712487},
publisher = {Association for Computing Machinery},
address = {New York, NY, USA},
url = {https://doi.org/10.1145/3691620.3695527},
doi = {10.1145/3691620.3695527},
booktitle = {Proceedings of the 39th IEEE/ACM International Conference on Automated Software Engineering},
pages = {1583–1594},
numpages = {12},
location = {Sacramento, CA, USA},
series = {ASE '24}
}

@inproceedings{10.1145/3643795.3648382,
author = {Piya, Sanyogita and Sullivan, Allison},
title = {LLM4TDD: Best Practices for Test Driven Development Using Large Language Models},
year = {2024},
isbn = {9798400705793},
publisher = {Association for Computing Machinery},
address = {New York, NY, USA},
url = {https://doi.org/10.1145/3643795.3648382},
doi = {10.1145/3643795.3648382},
booktitle = {Proceedings of the 1st International Workshop on Large Language Models for Code},
pages = {14–21},
numpages = {8},
location = {Lisbon, Portugal},
series = {LLM4Code '24}
}

@misc{pan2025hiddencostreadabilitycode,
      title={The Hidden Cost of Readability: How Code Formatting Silently Consumes Your LLM Budget}, 
      author={Dangfeng Pan and Zhensu Sun and Cenyuan Zhang and David Lo and Xiaoning Du},
      year={2025},
      eprint={2508.13666},
      archivePrefix={arXiv},
      primaryClass={cs.SE},
      url={https://arxiv.org/abs/2508.13666}, 
}

@misc{replicationPackage,
  author       = {Anonymous, Anonymous},
  title        = {Greening AI-Assisted Code Generation by Reducing
                   Babbling
                  },
  month        = mar,
  year         = 2026,
  publisher    = {Zenodo},
  doi          = {10.5281/zenodo.19237762},
  url          = {https://doi.org/10.5281/zenodo.19237762},
}


\end{document}